\documentclass[preprint]{my-elsarticle}

\usepackage{lineno,hyperref}
\usepackage{enumitem}
\usepackage{tikz}
\usepackage{amssymb,amsfonts,amsmath,mathtools,amsthm}
\usepackage{float}
\usepackage{graphicx}
\usepackage{resizegather}
\usepackage{longtable}
\usepackage{url}
\usepackage[numbers]{natbib}

\newtheorem{theorem}{Theorem}
\newtheorem{lemma}[theorem]{Lemma}

\usepackage{multirow,cases,subfigure}

\modulolinenumbers[5]

\journal{This is a preprint of a paper published 
in \emph{Franklin Open} at 
\url{https://doi.org/10.1016/j.fraope.2025.100397}}

\begin{document}

\begin{frontmatter}

\title{Evaluating the effectiveness of Stochastic CTMC and deterministic models in correlating rabies persistence in human and dog populations}

\author[mysecondaryaddress,mymainaddress]{Mfano Charles}\corref{mycorrespondingauthor}
\cortext[mycorrespondingauthor]{Corresponding author.}
\ead{mfanoc@nm-aist.ac.tz}

\author[comymainaddress]{Sayoki G. Mfinanga}
\ead{gsmfinanga@yahoo.com}

\author[mysecondaryaddress]{G.A. Lyakurwa}
\ead{geminpeter.lyakurwa@nm-aist.ac.tz}

\author[pt]{Delfim F. M. Torres}
\ead{delfim@ua.pt}

\author[mysecondaryaddress]{Verdiana G. Masanja}
\ead{verdiana.masanja@nm-aist.ac.tz}

\address[mysecondaryaddress]{School of Computational and Communication Science and Engineering,\\ 
Nelson Mandela African Institution of Science and Technology (NM-AIST), P.O. Box 447, Arusha, Tanzania}

\address[mymainaddress]{Department of ICT and Mathematics, 
College of Business Education (CBE), P.O. BOX 1968 Dar es Salaam, Tanzania}

\address[pt]{Center for Research and Development in Mathematics and Applications (CIDMA),\\
Department of Mathematics, University of Aveiro, 3810-193 Aveiro, Portugal}

\address[comymainaddress]{National Institute for Medical Research- Muhimbili Medical Research Centre, 
Box 3436, Dar es Salaam, Tanzania}


\begin{abstract}
Rabies continues to pose a significant zoonotic threat, particularly in areas 
with high populations of domestic dogs that serve as viral reservoirs. This 
study conducts a comparative analysis of Stochastic Continuous-Time Markov Chain (CTMC) 
and deterministic models to gain insights into rabies persistence within human and canine 
populations. By employing a multitype branching process, the stochastic threshold 
for rabies persistence was determined, revealing important insights into how stochasticity 
influences extinction probabilities. The stochastic model utilized 10,000 sample paths 
to estimate the probabilities of rabies outbreaks, offering a rigorous assessment 
of the variability in disease occurrences. Additionally, the study introduces a novel 
mathematical formulation of rabies transmission dynamics, which includes environmental 
reservoirs, free-ranging dogs, and domestic dogs as essential transmission factors. 
The basic reproduction number ($\mathcal{R}_0$) was derived and analyzed within 
stochastic frameworks, effectively bridging the gap between these two modeling 
approaches. Numerical simulations confirmed that the results from the stochastic 
model closely aligned with those from the deterministic model, while also highlighting 
the importance of stochasticity in scenarios with low infection rates. Ultimately, 
the study advocates for a comprehensive approach to rabies control that integrates 
both the predictable trends identified through deterministic models and the 
impact of random events emphasized by stochastic models.
\end{abstract}

\end{frontmatter}


\section{Introduction}

The ongoing presence of rabies among interconnected populations of dogs and humans poses a significant 
public health challenge, particularly in regions where access to medical resources is limited 
\cite{hueffer2024rabies}. Although rabies is preventable through vaccination, its continued 
prevalence in areas such as sub-Saharan Africa and parts of Asia underscores the persistent 
gaps in disease control \cite{haselbeck2021challenges}. 

One of the primary challenges in addressing rabies transmission in these regions is the 
underreporting of cases, particularly in rural and remote communities where healthcare 
infrastructure is minimal \cite{mbilo2021rabies}. This underreporting creates a critical 
gap in understanding the true extent of the disease, hindering effective surveillance, 
prevention, and intervention strategies. The situation is further complicated by environmental 
factors such as climate change and seasonality, which can exacerbate the spread of rabies 
and affect the interaction rates between dogs and humans. Variations in temperature and 
precipitation patterns can influence the behavior of rabies-carrying wildlife and 
domestic animals, contributing to unpredictable outbreaks. Additionally, the dynamics 
of dog populations in specific regions where free-roaming dogs are common and may not 
receive regular vaccinations serve as a significant reservoir for the virus, 
complicating efforts to control its transmission 
\cite{CHARLES2024115633,charles2024mathematical}.

Understanding the persistence of rabies in dog populations, along with the associated 
risks of human infection, necessitates the integration of both stochastic and 
deterministic modeling frameworks \cite{rysava2020towards}. A key strength 
of Stochastic Continuous-Time Markov Chain (CTMC) models is their ability 
to capture the inherent randomness of disease transmission, which is particularly 
relevant in contexts characterized by low numbers of infected individuals 
\cite{uchechukwu2023mathematical}. For any given compartment, each random 
event results in either the exit of an individual from the compartment 
or the entrance of an individual, unless the compartment is not involved. 
Therefore, for each process \( \mathbb{W}^{\theta}_i \), at a given 
time \( t \) such that \(\mathbb{W}^{\theta}_i(t) = w_i \in \mathbb{N} \), 
a transition is always of the form $w_i \to w_i + \varepsilon_i$, where
\begin{equation*}
\varepsilon_i =
\begin{cases}
-1 & \text{if an individual leaves compartment } i, \\
+1 & \text{if an individual enters compartment } i, \\
0 & \text{otherwise}.
\end{cases}
\end{equation*}
Consequently, the process \( \mathbb{W}^{\theta} \) is stochastic 
such that at each time \( t \), \( \mathbb{W}^{\theta}(t) \in \mathbb{N}^d \). 
This probabilistic framework makes Continuous-time Markov Chain (CTMC) models particularly 
effective for understanding the dynamics of rabies outbreaks and identifying the key factors 
contributing to persistent trends of the virus \cite{kouye2022sensitivity,curran2022stochastic}. 
In contrast, deterministic models offer a reliable approximation of overall disease dynamics 
and are instrumental in formulating large-scale intervention strategies, such as determining 
the vaccination coverage necessary to achieve herd immunity 
\cite{parino2023mathematical,duintjer2018using}. Nevertheless, deterministic models may fail 
to account for the influence of random events that can significantly alter disease dynamics, 
especially in situations where population sizes are small or where intervention efforts 
fluctuate over time \cite{duintjer2018using,faria2011toward,charles2025mathematical}. 
Despite these limitations, deterministic models remain essential tools for grasping 
overarching epidemiological trends and serve as a foundation for assessing the effectiveness 
of control measures. This comparison is vital for developing control strategies that address 
both predictable trends and the inherent variability of the transmission process, 
thereby ensuring a comprehensive approach to managing rabies outbreaks.

The structure of this paper is organized with 
sections dedicated to formulating the mathematical model 
(Section~\ref{modelFormulation}), qualitative analysis 
(Section~\ref{sec:02}), and quantitative analysis (Section~\ref{sec:03}).
We end with Section~\ref{sec:5} of discussion and  conclusion.


\section{Model Formulation}
\label{modelFormulation}

\subsection{A Deterministic Model}

A deterministic model describing the transmission dynamics of rabies among humans, 
free-range dogs, domestic dogs, and the environment 
is formulated, grounded on the following assumptions.

\begin{enumerate}[label=(\roman*)]
\item Rabies transmission occurs exclusively through effective contact 
between a susceptible host and an infectious host, or via contaminated 
environmental media (fomites, carcasses, or inanimate objects harboring the virus).

\item All infectious individuals are subject to both natural 
and disease-induced mortality, whereas non-infectious 
individuals experience only natural mortality.

\item Domestic dogs exhibit reduced susceptibility to infection 
due to human-provided protective measures, while free-range dogs 
receive neither PEP nor PrEP. Upon confirmed exposure, 
humans and domestic dogs receive effective PEP.

\item Recruitment rates in each population exceed corresponding natural 
mortality rates, ensuring persistence in the absence of disease, 
with recruitment assumed constant and unaffected by seasonal or stochastic variation.

\item Populations are homogeneously mixed, with uniform contact probabilities 
between individuals, regardless of spatial, social, or behavioral heterogeneity.

\item Humans and domestic dogs acquire temporary immunity following 
recovery, with immunity waning at a constant rate over time.
\end{enumerate}

\subsection*{Model Limitations}

\begin{enumerate}[label=(\roman*)]
\item Parameter values are assumed constant, neglecting  seasonal variability, 
and rare events such as outbreak fade-outs, which may limit realism in small populations.

\item Uniform contact rates overlook heterogeneity due to spatial segregation, 
social hierarchies, or human-mediated interactions, and the absence of age 
structure may omit key transmission dynamics.

\item Immunity decay is modeled as a constant-rate process, 
disregarding inter-individual variability in immune responses.

\item Environmental contamination is simplified, with limited representation 
of viral decay rates, persistence, or spatial clustering.

\item Assumes PEP is universally effective and accessible, which may 
not hold in resource-limited or rural settings.

\item Wildlife reservoirs and alternative host species are excluded, 
focusing solely on human--dog transmission dynamics.

\item The CTMC stochastic framework, although more representative of random processes, 
is computationally intensive, difficult to parameterize, and requires probability 
distributions that may not fully align with empirical data.
\end{enumerate}

\subsection{Description of Model Interaction} 

\begin{figure}[H]
\centering
\fbox{
\includegraphics[scale=0.7]{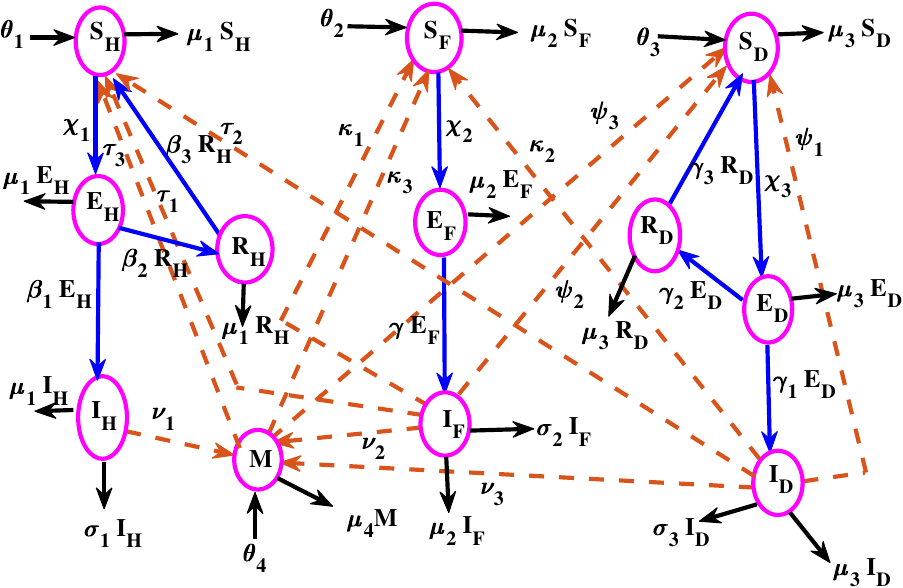}
}
\caption{\centering Flow diagram for rabies transmission among humans, 
free-range dogs, and domestic dogs.}
\label{fig:2}
\end{figure}

Susceptible humans are recruited at rate $\theta_{1}$ 
and become infected through contact with $I_{F}$, $I_{D}$, or the environment at 
rates $\tau_{1}$, $\tau_{2}$, and $\tau_{3}$, respectively. The infection rate is
\begin{equation*}
\chi_{1} = (\tau_{1} I_{F} + \tau_{2} I_{D} 
+ \tau_{3} \lambda(M)) S_{H}, \quad \lambda(M) = \dfrac{M}{M + C}.
\end{equation*}
Exposed humans ($E_{H}$) progress to $I_{H}$ at rate $\beta_{1}$ or recover 
with post-exposure prophylaxis at rate $\beta_{2}$. Immunity can wane at rate 
$\beta_{3}$, and the disease-induced death rate is $\sigma_{1}$. All human compartments 
experience a natural death rate of $\mu_{1}$. Free-range dogs are recruited at rate 
$\theta_{2}$ and become infected through contact with $I_{F}$, $I_{D}$, or the environment 
at rates $\kappa_{1}$, $\kappa_{2}$, and $\kappa_{3}$, respectively, with infection rate
\begin{equation*}
\chi_{2} = (\kappa_{1} I_{F} 
+ \kappa_{2} I_{D} + \kappa_{3} \lambda(M)) S_{F}.
\end{equation*}
Exposed free-range dogs ($E_{F}$) become $I_{F}$ at rate $\gamma$, 
with disease and natural death rates $\sigma_{2}$ and $\mu_{2}$. 
Domestic dogs are recruited at rate $\theta_{3}$ and infected 
at rates $\psi_{1}$, $\psi_{2}$, and $\psi_{3}$. Their infection rate is
\begin{equation*}
\chi_{3} = \left(\dfrac{\psi_{1} I_{F}}{1 + \rho_{1}} 
+ \dfrac{\psi_{2} I_{D}}{1 + \rho_{2}} 
+ \dfrac{\psi_{3}}{1 + \rho_{3}} \lambda(M)\right) S_{D}.
\end{equation*}
Exposed domestic dogs \(E_D\) progress to $I_{D}$ at rate $\gamma_{1}$ 
or recover at $\gamma_{2}$, with possible immunity loss at $\gamma_{3}$, 
disease-induced death at $\sigma_{3}$, and natural death at $\mu_{3}$. 
Virus shedding in the environment occurs from $I_{H}$, $I_{F}$, 
and $I_{D}$ at rates $\nu_1$, $\nu_2$, and $\nu_3$:
\begin{equation*}
\theta_{4} = (\nu_1 I_H + \nu_2 I_F + \nu_3 I_D) M,\; \text{with removal at rate}\;
 \mu_{4}.
\end{equation*}
The flow diagram presented in Figure~\ref{fig:2} illustrates the dynamics 
of rabies transmission, incorporating model assumptions, variable definitions, 
and parameter specifications.
By adopting a stochastic approach, the model captures the inherent randomness 
and variability in the transmission process, recognizing that real-world outcomes 
often deviate from average trends due to chance events. This extension is particularly 
useful for understanding the unpredictable nature of transmission, especially 
in smaller populations, where random fluctuations can result in unexpected outbreaks 
or even the elimination of the disease. A Deterministic Model of the rabies  
is described by system \eqref{eq:stochastic_model_with_noise} as
\begin{equation}
\left\{
\begin{aligned}
\dot{S_{H}}&=\theta_{1}+\beta_{3}R_{H}-\mu_{1} S_{H}-\chi_{1},\\
\dot{E_{H}}&=\chi_{1}-\left(\mu_{1}+\beta_{1}+\beta_{2}\right)E_{H},\\
\dot{I_{H}}&=\beta_{1}E_{H}-\left(\sigma_{1}+\mu_{1}\right) I_{H},\\
\dot{R_{H}}&=\beta_{2} E_{H}-\left(\beta_{3}+\mu_{1} \right) R_{H},\\\\
\dot{S_{F}}&=\theta_{2}-\chi_{2}-\mu_{2}S_{F},\\
\dot{E_{F}}&=\chi_{2}-\left(\mu_{2}+\gamma\right)E_{F},\\
\dot{I_{F}}&=\gamma E_{F}-\left(\mu_{2}+\sigma_{2}\right)I_{F},\\\\
\dot{S_{D}}&=\theta_{3}-\mu_{3}S_{D}-\chi_{3}+\gamma_{3}R_{D},\\
\dot{E_{D}}&=\chi_{3}-\left(\mu_{3}+\gamma_{1}+\gamma_{2}\right) E_{D},\\
\dot{I_{D}}&=\gamma_{1}E_{D}-\left(\mu_{3}+\sigma_{3}\right) I_{D},\\
\dot{R_{D}}&=\gamma_{2}E_{D}-\left(\mu_{3}+\gamma_{3}\right)R_{D},\\\\
\dot{M}&=\left(\nu_1I_H+\nu_2I_F+\nu_3I_D\right)-\mu_4M	.
\end{aligned}
\right.
\label{eq:stochastic_model_with_noise}
\end{equation}
subject to initial non-negative conditions  
\begin{equation*}
\begin{aligned}
&S_{H}(0) > 0, \; E_{H}(0) \geq 0, \; I_{H}(0) \geq 0, \; R_{H}(0) \geq 0, 
\; S_{F}(0) > 0, \; E_{F}(0) \geq 0, \; I_{F}(0) \geq 0, \\
&S_{D}(0) \geq 0, \; E_{D}(0) \geq 0, \; I_{D}(0) \geq 0, \; R_{D}(0) \geq 0. 
\end{aligned}
\end{equation*}


\section{Qualitative Analysis}
\label{sec:02}

In this section, we begin by proving the positivity and boundedness of the solutions of system
\eqref{eq:stochastic_model_with_noise} (Lemma~\ref{theorem:1} and Theorem~\ref{thm:bound}),
necessary conditions for the existence of a unique endemic equilibrium (Theorem~\ref{The}),
and global stability of the rabies disease-free  equilibrium point (Theorem~\ref{Th2}).
Then, we formulate a nonlinear continuous-time Markov chain (CTMC) stochastic model 
for rabies transmission dynamics and analyze its behavior employing
the theory of multitype branching processes near the rabies disease-free equilibrium point.


\subsection{Positivity of the Solutions and Boundedness of the System \eqref{eq:stochastic_model_with_noise}}
\label{sec:2.2.1}

We begin by proving existence and positivity.

\begin{lemma}
System \eqref{eq:stochastic_model_with_noise} admits a solution.
Moreover, all solutions of the system \eqref{eq:stochastic_model_with_noise} that  start in the region 
$\Omega \subset{\mathbb R }^{12}_{+}$ remain  positive all the time. 
\label{theorem:1}
\end{lemma}

\begin{proof}
To prove the existence of a solution to model \eqref{eq:stochastic_model_with_noise}, 
we consider initial  conditions  and apply the integral operator $\int\limits_{0}^{t} 
\left(\cdot\right)ds$  on each compartment of the model 
equation \eqref{eq:stochastic_model_with_noise} as
\begin{equation}
\dot{S}_H = \theta_1 + \beta_3 R_H - \mu_1 S_H 
- \left(\tau_1 I_F + \tau_2 I_D + \tau_3 \lambda(M)\right) S_H.
\label{p3}
\end{equation}
Integrating \eqref{p3} both sides over \([0, t]\), we get that
\begin{equation}
\displaystyle \int_0^t \dot{S}_H \, dt 
= \displaystyle \int_0^t \left(\theta_1 + \beta_3 R_H - \mu_1 S_H 
- \left(\tau_1 I_F + \tau_2 I_D + \tau_3 \lambda(M)\right) S_H\right) dt.
\label{p4} 
\end{equation}
Then, the  left-hand side of \eqref{p4} leads  to \(S_H(t) - S_H(0)\) 
and the right-hand side of the same equation leads to
\begin{equation}
\displaystyle \int_0^t \theta_1 \, dt + \displaystyle \int_0^t 
\beta_3 R_H \, dt - \displaystyle \int_0^t \mu_1 S_H \, dt 
- \displaystyle \int_0^t \left(\tau_1 I_F 
+ \tau_2 I_D + \tau_3 \lambda(M)\right) S_H \, dt.
\label{p5}
\end{equation}
By simplifying each integral in  \eqref{p5}, we have 
\begin{equation}
\displaystyle \int_0^t f(s) \, ds =
\begin{cases}
\theta_1 t, & \text{if } f(s) = \theta_1, \\
\beta_3 \displaystyle\int_0^t R_H(s) \, ds, 
& \text{if } f(s) = \beta_3 R_H(s), \\
\mu_1 \displaystyle\int_0^t S_H(s) \, ds, 
& \text{if } f(s) = \mu_1 S_H(s), \\
\displaystyle  \int_0^t \left(\tau_1 I_F(s) + \tau_2 I_D(s) 
+ \tau_3 \lambda(M)\right) S_H(s) \, ds, & \text{for disease dynamics}.
\end{cases}
\label{p6}
\end{equation}
By combining  and rearranging the  results in \eqref{p6}, it follows that
\begin{equation}
S_H(t) = S_H(0) + \theta_1 t + \beta_3 \int_0^t R_H(s) \, ds 
- \mu_1 \int_0^t S_H(s) \, ds - \int_0^t \left(\tau_1 I_F(s) 
+ \tau_2 I_D(s) + \tau_3 \lambda(M)\right) S_H(s) \, ds. 
\label{10}
\end{equation}
Thus, we conclude with the non-negativity of the integral terms
\begin{equation}
S_H(t)> S_H(0) + \theta_1 t + \beta_3 \int_0^t R_H(s) \, ds 
- \mu_1 \left(\tau_1 I_F(s) + \tau_2 I_D(s) + \tau_3 \lambda(M)\right) \int_0^t S_H(s) \, ds,
\end{equation}
since 
$$
S_H(0)> 0, \quad \theta_1 \geq 0, \quad \beta_3 R_H(s) \geq 0, \quad s \in [0, t].
$$
It follows  that $S_H(t) > 0$ for all $t \geq 0$. 
Similarly, we prove the  positivity of \( I_H(t) \) by considering
\begin{equation}
\dot{I}_{H} = \beta_{1}E_{H} - \left(\sigma_{1} + \mu_{1}\right) I_{H}.
\label{p1}
\end{equation}
Rearranging equation \eqref{p1} and applying the integrating factor \( \mu(t) \),  
it results in equation \eqref{p2}:
\begin{equation}
e^{(\sigma_{1} + \mu_{1}) t} I_{H} = I_{H}(0) 
+ \int_0^t \beta_{1} e^{(\sigma_{1} + \mu_{1}) s} E_{H}(s) ds.
\label{p2}
\end{equation}
By solving for \( I_H(t) \) in equation \eqref{p2}, we obtain 
\begin{equation*}
I_{H}(t) = e^{-(\sigma_{1} + \mu_{1}) t} \left( I_{H}(0) 
+ \int_0^t \beta_{1} e^{(\sigma_{1} + \mu_{1}) s} E_{H}(s) ds \right).
\end{equation*}
Since \( I_H(0) \geq 0 \) and \( E_H(s) \geq 0 \) for \( s \geq 0 \), 
it follows that
\begin{equation}
I_H(t) \geq 0 \quad \forall t \geq 0.
\end{equation}
Using the  same procedure, we conclude that 
\begin{equation*}
\begin{aligned}
&S_{H}(t) > 0, \; E_{H}(t) \geq 0, \; I_{H}(t) \geq 0, \; R_{H}(t) \geq 0, 
\; S_{F}(t) > 0, \; E_{F}(t) \geq 0, \; I_{F}(t) \geq 0, \\
&S_{D}(t) \geq 0, \; E_{D}(t) \geq 0, \; 
I_{D}(t) \geq 0, \; R_{D}(t) \geq 0, \forall t \geq 0,
\end{aligned}
\end{equation*}
and the proof is complete.
\end{proof}

\begin{theorem}
\label{thm:bound}
All solutions of system  \eqref{eq:stochastic_model_with_noise}  
starting in ${\mathbb{R}}^{12+}$ are uniformly bounded.
\end{theorem}

\begin{proof}
The model system  \eqref{eq:stochastic_model_with_noise}  
can be divided  in the subsection of human population, 
free range, and domestic dogs, as follows:
\begin{equation}
\begin{split}
\dfrac{d\left(S_H+E_H+I_H+R_H\right)}{dt} 
&= \theta_{1}+\beta_{3}R_{H}-\mu_{1} S_{H}-\left(\mu_{1}
+\beta_{1}+\beta_{2}+u_{4}\right)E_{H}
+\beta_{1}E_{H}  -\left(\sigma_{1}+\mu_{1}\right) I_{H}
+\beta_{2} E_{H}-\left(\beta_{3}+\mu_{1} \right) R_{H}.
\end{split}
\label{eqn83}
\end{equation}
Since the total number  of  human is given by \(S_H+E_H+I_H+R_H=N_{H}\), 
equation $\left(\ref{eqn83}\right)$  becomes
\begin{equation}
\dfrac{dN_H}{dt}= \theta_{1}-\left(S_{H}+E_{H}+I_{H}
+R_{H}\right)\mu_{1}-\sigma_{1}I_{H}.
\label{eqn82}
\end{equation}
We now recall the integrating  factor  on \eqref{eqn82}  as
\begin{eqnarray}
N_{H}\left(t\right)=e^{\displaystyle\int\limits_{0}^{t} \mu_{1}dt}
=e^{\mu_{1} t}
\label{13}
\end{eqnarray}
and, for $t\rightarrow 0$, equation$\left(\ref{13} \right)$ 
is simplified  as
\begin{eqnarray}
N_{H}(0)\le\dfrac{\theta_{1}}{\mu_{1}}+Ce^{0} 
\rightarrow N_{H}(0)-\dfrac{\theta_{1}}{\mu_{1}}\le C.
\label{16}
\end{eqnarray}
By simplifying  equation \eqref{16} with simple manipulation, we have
\begin{equation}
\Omega_{H} = \left\{ \left(S_{H}, E_{H}, I_{H}, R_{H}\right) 
\in \mathbb{R}_{+}^{4} : 0 \leq S_{H}+ E_{H}+I_{H}+R_H 
\leq \frac{\theta_{1}}{\mu_{1}} \right\}.
\end{equation}
So, using the same  procedure, it can be concluded that
\begin{align*}
\Omega_{F} &= \left\{ \left(S_{F}, E_{F}, I_{F}\right) 
\in \mathbb{R}_{+}^{3} : 0 \leq S_{F}+ E_{F}+ I_{F} 
\leq \frac{\theta_{2}}{\mu_{2}} \right\}, \\
\Omega_{D} &= \left\{ \left(S_{D}, E_{D}, I_{D}, R_{D}\right) 
\in \mathbb{R}_{+}^{4} : 0 \leq S_{D}+ E_{D}+ I_{D} 
+R_{D}\leq \frac{\theta_{3}}{\mu_{3}} \right\}, \\
M\left(t\right) &\leq \Omega_{M} = \max \left\{ \frac{\theta_1 
\nu_1}{\mu_1\mu_4}+\frac{\theta_2 \nu_2}{\mu_2\mu_4}
+\frac{\theta_3 \nu_3}{\mu_3\mu_4},M\left(0\right) \right\},
\end{align*}
and solutions are biologically and mathematically meaningfully: 
any solution relies in the region $\Omega$.
\end{proof}


\subsection{Rabies Persistent Equilibrium Point $E^{*}$} 
\label{sec:2.2.4}

The point $E^{*}$ denotes the steady-state condition at which rabies persists concurrently 
within the human population, free-range dog population, and domestic dog population. This 
equilibrium is determined by setting the right-hand sides of the governing equations 
in system~\eqref{eq:stochastic_model_with_noise} to zero and solving the resulting system 
of nonlinear equations simultaneously. The endemic equilibrium state is expressed as  
\begin{equation*}
E^{*}\left(S_{H}^{*},\; E_{H}^{*},\; I_{H}^{*},\; 
R_{H}^{*},\; S_{F}^{*},\; E_{F}^{*},\; I_{F}^{*},\; S_{D}^{*},\; 
E_{D}^{*},\; I_{D}^{*},\;  R_{D}^{*},\; M^{*}\right),
\end{equation*}  
where the components are given by  
\begin{equation*}
\left.
\begin{aligned}
R^{*}_H &= {\dfrac {\beta_{2} (\sigma_{1}+\mu_{1}) I^{*}_{H}}{\beta_{1} (\beta_{3}+\mu_{1})}}, \\\\
I^{*}_{H} &= \dfrac{\beta_1(\beta_3 + \mu_3)(\sigma_1 + \mu_1)^2(\beta_1 + \beta_2 
+ \beta_3)\mu_1 + \beta_1\beta_3(\sigma_1 + \mu_1)^2}{(\sigma_1 + \mu_1)^2((\beta_1 
+ \beta_2 + \beta_3)\mu_1 + \beta_1\beta_3)} \\
&\quad - \dfrac{\beta_1(\beta_3 + \mu_3)(\sigma_1 + \mu_1)^2\beta_3 
- \theta_1(\beta_3 + \mu_3)(\sigma_1 + \mu_1)^2}{(\sigma_1 
+ \mu_1)^2((\beta_1 + \beta_2 + \beta_3)\mu_1 + \beta_1\beta_3)}, \\\\
E^{*}_H &= {\dfrac {(\sigma_{1}+\mu_{1}) I^{*}_{H}}{\beta_{1}}}, \\
S^{*}_{H} &= \dfrac{\beta_{3}\beta_{2} (\sigma_{1}
+\mu_{1}) I^{*}_{H}}{\beta_{1} (\beta_{3}+\mu_{1}) \mu_{1}}
- \dfrac{(\mu_{1}+\beta_{1}+\beta_{2}) (\sigma_{1}+\mu_{1}) 
I^{*}_{H}}{\beta_{1}\mu_{1}}
+ \dfrac{\theta_{1}}{\mu_{1}}.
\end{aligned}
\right\}
\end{equation*}

\begin{equation*}
\left.
\begin{aligned}
I^{*}_{D} &= \dfrac{\gamma_{1}\psi_{1}I^{*}_{F}(1+\rho_{2})(1+\rho_{3})M^{*} 
+ \gamma_{1}\psi_{3}M^{*}(1+\rho_{1})(1+\rho_{2})}{(\mu_{3} + \gamma_{1} 
+ \gamma_{2})^2 - \gamma_{1}\psi_{2}(1+\rho_{1})(1+\rho_{3})M^{*}(\mu_{3} 
+ \gamma_{1} + \gamma_{2})}, \\\\
E^{*}_{D} &= {\dfrac {(\mu_{3}+\sigma_{3}) I^{*}_{D}}{\gamma_{1}}}, \quad
R^{*}_{D} = {\dfrac {\gamma_{2} (\mu_{3}+\sigma_{3}) I^{*}_{D}}{\gamma_{1} (\mu_{3}+\gamma_{3})}}.
\end{aligned}
\right\}
\end{equation*}

\begin{equation*}
\left.
\begin{aligned}
S^{*}_{D} &= \dfrac{\gamma_{3} (\mu_{3}+\sigma_{3})I^{*}_{D}}{\mu_{3} \gamma_{1}}
- \dfrac{(\mu_{3}+\gamma_{1}+\gamma_{2}) \gamma_{2} (\mu_{3}
+\sigma_{3}) I^{*}_{D}}{\gamma_{1} (\mu_{3}+\gamma_{3}) \mu_{3}}
+ \dfrac{\theta_{3}}{\mu_{3}}, \\
E^{*}_{F} &= {\dfrac {(\mu_{2}+\sigma_{2}) I^{*}_{F}}{\gamma}}, \\
S^{*}_{F} &= \dfrac{\theta_{2}}{\mu_{2}}
- \dfrac{(\mu_{2}+\gamma) (\mu_{2}+\sigma_{2}) I^{*}_{F}}{\gamma\,\mu_{2}}, \\
M^{*} &= {\dfrac {\nu_{3}I^{*}_{D}+\nu_{2}I^{*}_{F}+\nu_{1}I^{*}_{H}}{\mu_{4}}}.
\end{aligned}
\right\}
\end{equation*}

Here, the auxiliary parameters $\theta_{2}$ and $\theta_{3}$ are given by  
\begin{multline*}
\left.
\begin{aligned}
\theta_2 &= \dfrac{(\mu_2 + \gamma) \mu_2 (1 + (R_0 - 1)) (1 + \rho_1) \mu_3 
(\mu_2 + \sigma_2) \left( (1 + \rho_2)(\mu_3 + \sigma_3) (\mu_3 + \gamma_1 
+ \gamma_2) (1 + (R_0 - 1)) - \theta_3 \psi_2 \gamma_1 \right)}{\left( \mu_3 
(1 + \rho_2) (1 + \rho_1) (\mu_3 + \sigma_3) (\mu_3 + \gamma_1 + \gamma_2) 
(1 + (R_0 - 1)) - \theta_3 \gamma_1 (\psi_2 (1 + \rho_1) \mu_3 + \psi_1 
(1 + \rho_2)) \right) \gamma \kappa_1}, \\\\
\theta_3 &= \dfrac{\left( -\mu_2 (\mu_2 + \sigma_2) (\mu_2 + \gamma) (1 + (R_0 - 1)) 
+ \gamma \kappa_1 \theta_2 \right) (1 + (R_0 - 1)) (1 + \rho_1) \mu_3 (1 + \rho_2) 
(\mu_3 + \sigma_3) (\mu_3 + \gamma_1 + \gamma_2)}{\left( \left( -\mu_2 (\mu_2 + \sigma_2) 
(\mu_2 + \gamma) (1 + (R_0 - 1)) + \gamma \kappa_1 \theta_2 \right) (1 + \rho_1) \psi_2 \mu_3 
+ \gamma \kappa_1 \theta_2 \psi_1 (1 + \rho_2) \right) \gamma_1}.
\end{aligned}
\right\}
\end{multline*}

The endemic equilibrium exists when $I_{H} > 0$, $I_{F} > 0$, $I_{D} > 0$, $M > 0$, 
and ${\mathcal R}_0 \geq 1$, as summarized in Theorem~\ref{The}.

\begin{theorem}
The model system~\eqref{eq:stochastic_model_with_noise} possesses a unique endemic 
equilibrium $E^{*}$ if $\mathcal{R}_0 \geq 1$ and $E_{H}, E_{F}, E_{D}, M > 0$.
\label{The}
\end{theorem}

The proof that the endemic equilibrium point  $E^{*}$ of the rabies model 
\eqref{eq:stochastic_model_with_noise} is globally asymptotically stable 
whenever $\mathcal{R}_{0} \ge 1$ is given in \ref{Appendix:A}.


\subsection{Global Stability of the Rabies Disease Free Equilibrium Point \(E^{0}\)}
\label{subsec:GS:DFE}

To obtain $E^{0}$, the  left hand  side  of  equation in the model system  
\eqref{eq:stochastic_model_with_noise}  is set to zero, such that
\begin{align*}
E^{0} 
&= \left(\begin{array}{cccccccccccc}
\dfrac{\theta_{1}}{\mu_{1}}, & 0, & 0, & 0, 
& \dfrac{\theta_{2}}{\mu_{2}}, & 0, & 0, 
& \dfrac{\theta_{3}}{\mu_{3}}, & 0, & 0, & 0, & 0
\end{array}\right).
\end{align*}

\begin{theorem}
\label{Th2}	
The rabies disease  free equilibrium point \(E^{0}\) is  globally  
asymptotically stable  when \(\mathcal{R}_{0}<0\) and unstable otherwise.
\end{theorem}

\begin{proof}
The analysis of the equilibrium behavior  
\(E^{0}\) of the model described in \eqref{eq:stochastic_model_with_noise} 
employs the Metzler matrix, as demonstrated by \cite{castillo2002computation} 
and \cite{charles2024mathematical}. In this  context, $U_{s}$ represent the 
compartments that do not transmit rabies, and $U_{i}$ represent the rabies-transmitting 
compartments. If $A_{2}$ is a Metzler matrix (with non-negative off-diagonal entries) 
and $A_{0}$ has real negative eigenvalues, then the rabies-free equilibrium is globally 
asymptotically stable. Then, the model equation \eqref{eq:stochastic_model_with_noise} 
is decomposed  to
\begin{equation}
\begin{cases}
\begin{array}{llll}
\dfrac{dU_{s}}{dt} 
&= A_{0}\left(U_{s}-U\left(E^{0}\right)\right)+A_{1}U_{i}, \\
\dfrac{dU_{i}}{dt} 
&= A_{2}U_{i},
\end{array}
\end{cases}
\end{equation}
where
\begin{equation*}
U_{s} - U\left( E^{0}\right)
= \left(
\begin{array}{c}
S_{H} - \dfrac{\theta_{1}}{\mu_{1}} \\[8pt]
R_{H} \\[8pt]
S_{F} - \dfrac{\theta_{2}}{\mu_{2}} \\[8pt]
S_{D} - \dfrac{\theta_{3}}{\mu_{3}} \\[8pt]
R_{D}
\end{array}
\right),
\quad
A_{0} =
\left(
\begin{array}{ccccc}
-\mu & \beta_{3} & 0 & 0 & 0 \\[8pt]
0 & -\left(\beta_{3} + \mu_{1}\right) & 0 & 0 & 0 \\[8pt]
0 & 0 & -\mu_{2} & 0 & 0 \\[8pt]
0 & 0 & 0 & -\mu_{3} & \gamma_{3} \\[8pt]
0 & 0 & 0 & 0 & -\left(\mu_{3} + \gamma_{3}\right),
\end{array}
\right),
\end{equation*}
\begin{equation*}
A_{1}=
\left(
\begin {array}{ccccccc} 
0&0&0&{\dfrac {\tau_{{1}}\theta_{{1}}}{
\mu_{{1}}}}&0&{\dfrac {\tau_{{2}}\theta_{{1}}}{\mu_{{1}}}}&0\\ 
\noalign{\medskip}\beta_{{2}}&0&0&0&0&0&0\cr \noalign{\medskip}0&0&0
&{\frac {\kappa_{{1}}\theta_{{2}}}{\mu_{{2}}}}&0&{\dfrac {\kappa_{{2}}
\theta_{{2}}}{\mu_{{2}}}}&0\cr \noalign{\medskip}0&0&0&{\dfrac {\psi_{{1}}
\theta_{{3}}}{\mu_{{3}} \left( 1+\rho_{{1}} \right) }}&0
&{\dfrac {\psi_{{2}}\theta_{{3}}}{\mu_{{3}} \left( 1+\rho_{{2}} \right) }}
&0\cr \noalign{\medskip}0&0&0&0&\gamma_{{2}}&0&0
\end{array}
\right),
\end{equation*}
\begin{equation*}
\text{and}\;\;A_{2} =	
\left(
\begin {array}{ccccccc} 	
-\mu_{{1}}-\beta_{{1}}-\beta_{{2}}&0&0
&{\frac {\tau_{{1}}\theta_{{1}}}{\mu_{{1}}}}&0&{\frac {\tau_{{2}}
\theta_{{1}}}{\mu_{{1}}}}&0\\ \noalign{\medskip}\beta_{{1}}
&-\sigma_{{1}}-\mu_{{1}}&0&0&0&0&0\\ \noalign{\medskip}0&0
&-\mu_{{2}}-\gamma
&{\frac {\kappa_{{1}}\theta_{{2}}}{\mu_{{2}}}}&0
&{\frac {\kappa_{{1}}\theta_{{2}}}{\mu_{{2}}}}&0\\ 
\noalign{\medskip}0&0&\gamma&-\mu_{{2}}-
\sigma_{{2}}&0&0&0\\ \noalign{\medskip}0&0&0&{\frac {\psi_{{1}}
\theta_{{3}}}{\mu_{{3}} \left( 1+\rho_{{1}} \right) }}
&-\mu_{{3}}-\gamma_{{1}}-\gamma_{{2}}
&{\frac {\psi_{{2}}\theta_{{3}}}{\mu_{{3}} \left( 
1+\rho_{{1}} \right) }}&0\\ \noalign{\medskip}0&0&0
&0&\gamma&-\mu_{{3}}-\sigma_{{3}}&0\\ \noalign{\medskip}0
&\nu_{{1}}&0& \nu_{{2}}&0&\nu_{{3}}&-\mu_{{4}}
\end{array} \right). 
\end{equation*}
Given that the eigenvalues of the matrix \(A_{0}\) are negative 
and the off-diagonal entries of the Metzler matrix \(A_{2}\) 
are non-negative, it follows that the rabies equilibrium 
point \(E^{0}\) is globally asymptotically stable. 
\end{proof}


\subsection{Rabies CTMC Stochastic Model Formulation}
\label{sec:CTMC}

Continuous Time Markov Chain (CTMC) stochastic models utilize Galton--Watson 
Branching Processes to delineate the probabilities of various events, 
offering valuable insights into dynamics, control strategies, prediction 
of expected case numbers, extinction time, and the assessment of surveillance 
and response system effectiveness. While deterministic models rely on the basic 
reproduction number $\left(\mathcal{R}_0\right)$ to determine whether a disease 
persists or diminishes in a population, stochastic models view $\mathcal{R}_0$ 
as a stochastic threshold, recognizing that the disease can still cease to exist 
even if the threshold exceeds one, contingent upon the initial number of 
infectives introduced into a susceptible population. 

A continuous time Markov chain (CTMC) stochastic model for rabies transmission dynamics  
has been developed based on the assumptions employed in the stochastic model  
\eqref{eq:stochastic_model_with_noise}. For the sake of simplicity, the same 
notations and parameters as those used in the deterministic model have been adopted. 
Let $S_H, E_H, I_H, R_H, S_F, E_F, I_F, S_D, E_D, I_D, R_D, M$, denote the discrete 
random  variable  for susceptible humans, exposed humans, infectious humans, 
recovered humans, susceptible free-range dogs, exposed free-range dogs, 
infectious free-range dogs, susceptible domestic dogs, exposed domestic dogs, 
infectious domestic animals, recovered domestic  dogs, and environment respectively.
Let
\begin{equation*}
\mathbf{X}\left(t\right) 
= \left(S_H, E_H, I_H, R_H, S_F, E_F, I_F, S_D, E_D, I_D, R_D, M\right)^{T}
\end{equation*}
be the associated random vector for all discrete random
variables  $S_H$, $E_H$, $I_H$, $R_H$, $S_F$, $E_F$, $I_F$, 
$S_D$, $E_D$, $I_D$, $R_D$, and $M$. 
Given the time-homogeneous nature of the Continuous-Time Markov Chain (CTMC) 
model and its adherence to the Markov property, it is established that the 
future state of the process at \((t + \Delta t)\) hinges entirely upon 
the current state at time \(t\). As a result, the interval between events 
follows an exponential distribution characterized by a specific parameter:
\begin{equation}
\begin{cases}
\Psi\left(\textbf{X}\right) 
&= \theta_1 + \beta_3 R_H + \mu_1 N_H + \tau_1 I_F S_H 
+ \tau_2 I_D S_H + \left( \dfrac{\tau_3 M}{M + C}\right) S_H + \beta_1 E_H \\
&\quad + \kappa_1 I_F S_F + \kappa_2 I_D S_F + \left( 
\dfrac{\kappa_3 M}{M + C}\right) S_F + \mu_2 N_F + \gamma E_F + \sigma_2 I_F + \theta_3 \\
&\quad + \left(\dfrac{\psi_1}{1 + \rho_1}\right) I_F S_D 
+ \left(\dfrac{\psi_2}{1 + \rho_2}\right) I_D S_D 
+ \left(\frac{\psi_3 M}{\left(1 + \rho_3\right) \left(M + C\right)}\right) S_D \\
&\quad + \mu_3 N_D + \gamma_3 R_D + \gamma_1 E_D + \gamma_2 E_D 
+ \sigma_3 I_D + \nu_1 I_H + \nu_2 I_F + \nu_3 I_D \\
&\quad + \mu_4 M + \beta_2 E_H + \sigma_1 I_H + \theta_2,
\end{cases}
\end{equation}
where
\begin{equation*}
N_H=S_H+E_H+I_H+R_H, \; N_F=S_F+E_F+I_F \; 
\textup{and}\; N_D=S_D+E_D+I_D+R_D.
\end{equation*}
For modeling the transmission dynamics of rabies between humans and dogs 
using a Continuous Time Markov Chain (CTMC) model, event transitions 
and their corresponding rates are typically derived from the deterministic model. 
These transitions occur as individuals move between compartments due to recruitment 
or movement, assuming an initial presence of only one individual while other 
sub-populations are not yet established. Table~\ref{tab:transitions} summarizes 
the events and their associated transition rates, where  the values 1, -1, and 0 
represent an increase by 1, a decrease by 1, and no change in state, respectively, 
for the variable from time \( t \) to \( t + \Delta t \).
\begin{center}
\begin{longtable}{|l|l|l|}
\caption{State transitions and rates of occurrence for the CTMC.} \\
\hline 
\textbf{Event} & \textbf{Rate, r} & \textbf{Transition $\Delta \tilde{Z}(t)$} \\
\hline\hline
\endfirsthead
\multicolumn{3}{c}
{\tablename\thetable\ -- \textit{Continued from previous page}} \\
\hline
\textbf{Event} & \textbf{Rate, r} & \textbf{Transition $\Delta \tilde{Z}(t)$} \\
\hline
\endhead
\hline \multicolumn{3}{r}{\textit{Continued on next page}} \\
\endfoot
\hline
\endlastfoot
\hline
Recruitment of $S_H$ & $\theta_1$ &  $\left( 1, 0, 0, 0, 0, 0, 0, 0, 0, 0, 0, 0\right)$\\ \hline
Natural death  of $S_H$  & $\mu_1 S_H$ &  $\left( - 1, 0, 0, 0, 0, 0, 0, 0, 0, 0, 0, 0\right) $ \\ \hline
Contact   of  $S_H$ and  $I_F$ &$\tau_1 I_F S_H$  & $ \left(- 1, 1, 0, 0, 0, 0, 0, 0, 0, 0, 0, 0\right)$  \\ \hline
Contact  of  $S_H$ and  $I_D$ &$\tau_2 I_DS_H$  & $\left(- 1, 1, 0, 0, 0, 0, 0, 0, 0, 0, 0, 0\right)$ \\ \hline
Contact   of  $S_H$ and  $M$ &$\dfrac{M\tau_3}{M+C}S_H $  & $\left(- 1, 1, 0, 0, 0, 0, 0, 0, 0, 0, 0, 0\right)$ \\ \hline
Progression  from $E_H$  to  $I_H$ &$\beta_1 E_H$& $\left(0, -1, 1, 0, 0, 0, 0, 0, 0, 0, 0, 0\right)$ \\ \hline
Recovery   of $E_H$ & $\beta_2 E_H$ & $\left(0, -1, 0, 1, 0, 0, 0, 0, 0, 0, 0, 0\right)$  \\ \hline
Natural death of   $E_H$ & $ \mu_1 E_H$  & $\left(0, -1, 0, 0, 0, 0, 0, 0, 0, 0, 0, 0\right)$  \\ \hline
Disease induced death of  $I_H$ & $ \sigma_1 I_H$  & $\left(0, 0, -1, 0, 0, 0, 0, 0, 0, 0, 0, 0\right)$  \\ \hline
natural death of  $I_H$ & $ \mu_1 I_H$  & $\left(0, 0, -1, 0, 0, 0, 0, 0, 0, 0, 0, 0\right)$  \\ \hline
Natural death  of $R_H$&$\mu_1 R_H$ &$\left(0, 0, 0, -1, 0, 0, 0, 0, 0, 0, 0, 0\right)$   \\ \hline
Rate of immunity  loss of $R_H$&$\beta_3 R_H$ &$\left(1, 0, 0, -1, 0, 0, 0, 0, 0, 0, 0, 0\right)$\\ \hline
Recruitment  of  $S_F $& $\theta_2$ & $\left(0, 0, 0, 0, 1, 0, 0, 0, 0, 0, 0, 0\right)$ \\ \hline
Natural death  of $S_F $ & $\mu_2 S_F$ & $\left(0, 0, 0, 0, -1, 0, 0, 0, 0, 0, 0, 0\right)$   \\ \hline
Contact of  $S_F$ and  $I_F$ & $\kappa_1 I_F S_F $ & $\left(0, 0, 0, 0, -1, 1, 0, 0, 0, 0, 0, 0\right)$ \\ \hline
Contact of  $S_F$ and  $I_D$ & $\kappa_2 I_D S_F $ & $\left(0, 0, 0, 0, -1, 1, 0, 0, 0, 0, 0, 0\right)$\\ \hline
Contact of  $S_F$ and  $M$ & $\dfrac{M\kappa_3}{M+C}  S_F $ & $\left(0, 0, 0, 0, -1, 1, 0, 0, 0, 0, 0, 0\right)$ \\ \hline
Progression of $E_F$ to $ I_F$ & $\gamma E_F$ & $\left(0, 0, 0, 0, 0, -1, 1, 0, 0, 0, 0, 0\right)$  \\ \hline
Natural death  of $E_F$  & $\mu_2 E_F$ & $\left(0, 0, 0, 0, 0, -1, 0, 0, 0, 0, 0, 0\right)$  \\ \hline
Natural death  of $I_F$  & $\mu_2 I_F$ & $\left(0, 0, 0, 0, 0, 0, -1, 0, 0, 0, 0, 0\right)$   \\ \hline
Disease induced death of $I_F$ & $\sigma_2 I_F$ & $\left(0, 0, 0, 0, 0, 0, -1, 0, 0, 0, 0, 0\right)$ \\ \hline
Recruitment of $S_D$ & $\theta_3$ & $\left(0, 0, 0, 0, 0, 0, 0, 1, 0, 0, 0, 0\right)$ \\ \hline
Natura death of $S_D$ & $\mu_3 S_D$  &  $\left(0, 0, 0, 0, 0, 0, 0, -1, 0, 0, 0, 0\right)$\\ \hline
Contact of  $S_D$ and  $I_F$ & $\dfrac{\psi_1 I_F }{1+\rho_{1}}  S_D $ 
& $\left(0, 0, 0, 0, 0, 0, 0, -1, 1, 0, 0, 0\right)$ \\ \hline
Contact of  $S_D$ and  $I_D$ & $\dfrac{\psi_2 I_D }{1+\rho_{2}}  S_D $ 
& $\left(0, 0, 0, 0, 0, 0, 0, -1, 1, 0, 0, 0\right)$ \\ \hline
Contact of  $S_D$ and  $M$ & $\dfrac{\psi_3 M}{\left(1+\rho_{3}\right)\left(C+M\right)}  S_D$ 
& $\left(0, 0, 0, 0, 0, 0, 0, -1, 1, 0, 0, 0\right)$ \\ \hline
Progression  of $E_D$  to $I_D$  & $\gamma_1 E_D$  & $\left(0, 0, 0, 0, 0, 0, 0, 0, -1, 1, 0, 0\right)$\\ \hline
Recovery of $E_D$  & $\gamma_2 E_D$& $\left(0, 0, 0, 0, 0, 0, 0, 0, -1, 0, 1, 0\right)$  \\ \hline
Disease induced death of  $I_D$ &$\sigma_3 I_D$ &  $\left(0, 0, 0, 0, 0, 0, 0, 0, 0 -1, 0, 0\right)$ \\ \hline
Natural death  of $E_D$ & $ \mu_3 E_D$ & $\left(0, 0, 0, 0, 0, 0, 0, 0, -1 0, 0, 0\right)$  \\ \hline
Natural death  of $I_D$ & $ \gamma_3 R_D$ & $\left(0, 0, 0, 0, 0, 0, 0, 0, 0 -1, 0, 0\right)$  \\ \hline
Natural death  of $R_D$ & $ \mu_3 R_D$ & $\left(0, 0, 0, 0, 0, 0, 0, 0, 0, 0, -1, 0\right)$  \\ \hline
Remove  of  rabies in the environment & $\mu_4 M$ & $\left(0, 0, 0, 0, 0, 0, 0, 0, 0, 0, 0, -1\right)$ \\ \hline
shading  of $I_H$  to $M$ &$\nu_1 I_H$ &$\left(0, 0, 0, 0, 0, 0, 0, 0, 0, 0, 0, 1\right)$ \\ \hline
shading  of $I_F$  to $M$ &$\nu_2 I_F$ & $\left(0, 0, 0, 0, 0, 0, 0, 0, 0, 0, 0, 1\right)$\\ \hline
shading  of $I_D$  to $M$ &$\nu_3 I_D$ & $\left(0, 0, 0, 0, 0, 0, 0, 0, 0, 0, 0, 1\right)$ 
\label{tab:transitions}
\end{longtable}
\end{center}


\subsection{Multitype Branching Process}

The multitype branching process theory is employed to analyze the behavior 
of the nonlinear Continuous-Time Markov Chain (CTMC) near the Rabies-free 
equilibrium point ${\mathbb E}_0$. This theory is crucial for determining 
the probabilities of disease extinction or outbreak under various initial 
conditions. In CTMC models, the branching process can either grow exponentially 
or diminish to zero, particularly when the initial number of infectives 
is minimal at the onset of a disease outbreak. In order for the multitype 
branching process to be applicable, it necessitates a sufficiently large 
initial susceptible population. As per the parameters of this study, 
we have established the initial susceptible populations as follows:
\(S_H\left(0\right)=\dfrac{\theta_1}{\mu_{1}}\),
\(S_F\left(0\right)=\dfrac{\theta_2}{\mu_{2}}\), 
and \(S_D\left(0\right)=\dfrac{\theta_3}{\mu_{3}}\). 
We assume that infectives of type \(i,\; I_i\), produce infectives 
of type \(j,\;I_j\), and the number of offspring produced by an 
individual of type \(I_i\) is independent of the number 
of offspring produced by either type \(I_i\) or type \(I_j\), 
where \(j \neq i\). The term 'birth' describes the infection 
transmission between susceptible humans, infected humans, susceptible 
domestic dogs, infected domestic dogs, susceptible free range dogs, 
infected free range dogs, and the  rabies in the environment. Since 
the multitype branching process is linear near the disease-free equilibrium, 
the numbers of deaths and births are independent. We define probability 
generating functions (pgfs) for the births and deaths of rabies in the  
environment, infected humans, domestic and free range dogs, which are essential 
for determining the probability of rabies  extinction or outbreak in humans and dogs.

Let \(\{Y_{ji}\}_{j=1}^n\) be the offspring random variable for type \(i\), 
where \(i = 1, 2, \ldots, n\) infectious hosts. Here, \(Y_{ji}\) represents 
the number of offspring of type \(j\) produced by an infective of type \(i\). 
The offspring probability generating function (pgf) for the infectious population 
\(I_i\) is defined under the condition that there is initially one infectious host 
at the beginning of the disease outbreak, i.e., \(I_i(0) = 1\), and all other types 
are zero: \(I_j = 0\). The offspring pgf \(f_i : [0,1]^n \to [0,1]\) for type \(i\) 
individuals, given \(I_i(0) = 1\) and \(I_j(0) = 0\) for \(j \neq i\), is expressed as
\begin{equation}
f_i(u_1, u_2, \ldots, u_n) = \sum_{\ell_1=0}^{\infty} 
\sum_{k_2=0}^{\infty} \cdots \sum_{\ell_n=0}^{\infty} 
P_i(\ell_1, \ell_2, \ldots, \ell_n) u_1^{\ell_1} u_2^{\ell_2} \cdots u_n^{\ell_n},
\label{prob1}
\end{equation}
where
\begin{equation}
P_i(\ell_1, \ell_2, \ldots, \ell_n) = \text{Prob}\{Y_{1j} 
= \ell_1, Y_{2j} = \ell_2, \ldots, Y_{nj} = \ell_n\}
\end{equation}
is the probability that a single infectious individual of type $i$ 
will produce $k$ offspring of type $j$. Equation \eqref{prob1} is utilized 
to establish an $n \times n$ non-negative and irreducible expectation matrix 
$\mathcal{M}_1 = [m_{ji}]$, where $m_{ji}$ denotes the expected number 
of offspring of type $j$ generated by an infected individual of type $i$. 
The elements of matrix $\mathcal{M}_1$ are calculated by differentiating 
$f_i$ with respect to $u_j$ and then evaluating all $u$ variables at 1 
\cite{lahodny2015estimating, maity2022comparison}, meaning that
\begin{equation}
m_{ji} = \frac{\partial f_i}{\partial u_j}\Bigg|_{u=1} < \infty. 
\end{equation}
The probability of disease extinction or outbreak is determined 
by the size of the spectral radius of the expectation matrix 
$\mathcal{M}_1$, $\rho(\mathcal{M}_1)$. If $\rho(\mathcal{M}_1) \leq 1$, 
then the probability of disease extinction is one, that is,
\begin{equation}
\mathbb{P}_0 = \lim_{t \to \infty} 
\text{Prob}\{\tilde{I}(t) = \tilde{0}\} = 1,
\end{equation}
and if $\rho(\mathcal{M}_1) > 1$, then there exists a positive 
probability such that the probability of disease extinction is given by
\begin{equation}
\mathbb{P}_0 = \lim_{t \to \infty} \text{Prob}\{\tilde{I}(t) 
= \tilde{0}\} = q_1^{i_1} q_2^{i_2} \ldots q_k^{i_k} < 1, 
\end{equation}
where $(q_1, q_2, \ldots, q_k)$ is the unique fixed point of the $k$ 
offspring pgf, $f_i(q_1, q_2, \ldots, q_k) = q_i$, 
and $0 < q_i < 1$, $i = 1, 2, \ldots, k$ 
\cite{lahodny2015estimating,MR3078807,MR4561181}.
The probability of disease outbreak is 
\begin{equation}
1 - \mathbb{P}_0 = 1 - q_1^{i_1} q_2^{i_2} \ldots q_k^{i_k},
\end{equation}
where $\mathbb{P}_0 $ is the probability of  extinction or outbreak
\cite{lahodny2015estimating, maity2022comparison}. 
Predictions concerning disease extinction and the occurrence of outbreaks 
can be made using stochastic epidemic theory, which focuses on the number 
of infectious individuals within each group. If a disease originates from 
an infectious group with a reproduction number (\(\mathcal{R}_0>1\)), 
and \(i\) infective individuals are introduced into an entirely susceptible 
population, the probability of a significant outbreak is approximately 
\(1 - \left(\dfrac{1}{\mathcal{R}_0}\right)^i\). Conversely, the probability 
of the disease becoming extinct is approximately 
\(\left(\dfrac{1}{\mathcal{R}_0}\right)^i\). In the early stages 
of a rabies outbreak, with only a few infected dogs, there is a limited potential 
for generating infectious humans and dogs. Exposed humans and both free-range 
and domestic dogs can progress to infectious classes. The offspring probability 
generating function for the exposed class $E_H$, given that $E_H\left(0\right)=1$, 
$I_H\left(0\right)=0$, $E_F\left(0\right)=0$, $I_F\left(0\right)=0$, 
$E_D\left(0\right)=0$, $I_D\left(0\right)=0$, and $M\left(0\right)=0$, 
is given by
\begin{equation}
f_1\left(u_1\;,u_2\;,u_3\;,u_4,\;\ldots, u_7\;\right)
=\dfrac{\beta_1 u_2+\beta_2+\mu_1}{\beta_1+\beta_2+\mu_1}.
\label{pgf1}
\end{equation}
The expression ${\beta_1}\slash{\beta_1+\beta_2+\mu_1}$ denotes the 
probability of exposed individual progressing to the infectious class $I_H$. 
The term ${\beta_2}\slash{\beta_1+\beta_2+\mu_1}$ represents the probability 
of exposed individuals recovering as a result of Rabies Postexposure Prophylaxis (PEP), 
while ${\mu_1}\slash{\beta_1+\beta_2+\mu_1}$ indicates the probability of exposed 
individuals naturally dying  before transitioning to the infected class.

If $E_H\left(0\right)=0$, $I_H\left(0\right)=1$, $E_F\left(0\right)=0$, 
$I_F\left(0\right)=0$, $E_D\left(0\right)=0$, $I_D\left(0\right)=0$, 
and $M\left(0\right)=0$, then the offspring probability generating 
function for $I_H$ is given by
\begin{equation}
f_2\left(u_1\;,u_2\;,u_3\;,u_4,\;\ldots, u_7\;\right)
=\dfrac{\nu_1 u_2 u_7+\sigma_1+\mu_1}{\nu_1+\sigma_1+\mu_1}.
\label{pgf2}
\end{equation}
In pgf \eqref{pgf2}, the term ${\nu_1}\slash{\mu_1+\nu_1+\sigma_1}$ represents 
the probability of infected humans shedding the rabies virus in the environment;
${\mu_1}\slash{\mu_1+\nu_1+\sigma_1}$ signifies the probability of infected humans 
dying naturally; and the term ${\sigma_1}\slash{\mu_1+\nu_1+\sigma_1}$ refers 
to the probability of infected humans dying due to the disease.

The offspring probability generating function for $E_F$, such that
$E_H\left(0\right)=0$, $I_H\left(0\right)=0$, $E_F\left(0\right)=1$, 
$I_F\left(0\right)=0$, $E_D\left(0\right)=0$, $I_D\left(0\right)=0$, 
and $M\left(0\right)=0$, is  given by
\begin{equation}
f_3\left(u_1, u_2, u_3, u_4,\;\ldots, u_7\;\right)
=\dfrac{\gamma u_4+\mu_2}{\gamma+\mu_2},
\label{pgf3}
\end{equation}
where ${\gamma} \slash{\gamma+\mu_2}$ denotes the probability of the exposed  
free range dogs class progressing to the infected class,  
and ${\mu_2}\slash{\gamma+\mu_2}$ represents the probability 
of exposed free dogs dying naturally before progressing to the infected class. 

The offspring  probability for $I_F$, given that $E_H\left(0\right)=0$, 
$I_H\left(0\right)=0$, $E_F\left(0\right)=0$, $I_F\left(0\right)=1$, 
$E_D\left(0\right)=0$, $I_D\left(0\right)=0$, and $M\left(0\right)=0$, is given by
\begin{equation}
\begin{aligned}
f_4\left(u_1, u_2, u_3, u_4,\;\ldots, u_7\;\right)
=\dfrac{\hat{\lambda}_{1} u_1 u_4+\hat{\lambda}_{2} u_4 u_5
+\hat{\lambda}_{3} u_3 u_4+\mu_2+\sigma_2+\nu_2 u_4 u_7}{\hat{\lambda}_{1}
+\hat{\lambda}_{2}+\hat{\lambda}_{3}+\sigma_2+\mu_2+\nu_2},\;\;\\  
\textup{for}\; \hat{\lambda}_{1}=\tau_1 S^{0}_{H},\;\hat{\lambda}_{2}
=\dfrac{\psi_1}{1+\rho_1} S^{0}_{D},\; \hat{\lambda}_{3}=\kappa_1 S^{0}_{F}.
\end{aligned}
\label{pgf4}
\end{equation}
In pgf \eqref{pgf4}, the term  ${\hat{\lambda}_{1}}\slash{\hat{\lambda}_{1}
+\hat{\lambda}_{2}+\hat{\lambda}_{3}+\sigma_2+\mu_2+\nu_2}$ represents  
the probability  of infected free range  dogs  to cause  rabies infection  
to susceptible  humans, ${\hat{\lambda}_{2}}\slash{\hat{\lambda}_{1}
+\hat{\lambda}_{2}+\hat{\lambda}_{3}+\sigma_2+\mu_2+\nu_2}$ signifies 
the probability of an infected free range dogs  causing infection  
to  susceptible domestic dogs, ${\hat{\lambda}_{3}}\slash{\hat{\lambda}_{1}
+\hat{\lambda}_{2}+\hat{\lambda}_{3}+\sigma_2+\mu_2+\nu_2}$ denotes  
an infected free range dogs causing infection to free range dogs, 
${\nu_2}\slash{\hat{\lambda}_{1}+\hat{\lambda}_{2}+\hat{\lambda}_{3}
+\sigma_2+\mu_2+\nu_2}$  represents  the probability of free range dogs  
to shade rabies virus in the environment, ${\mu_2}\slash{\hat{\lambda}_{1}
+\hat{\lambda}_{2}+\hat{\lambda}_{3}+\sigma_2+\mu_2+\nu_2}$ denotes the 
probability of infected  free range  dogs  dying naturally,  and  the term 
${\sigma_2}\slash{\hat{\lambda}_{1}+\hat{\lambda}_{2}+\hat{\lambda}_{3}
+\sigma_2+\mu_2+\nu_2}$ is the probability of infected free range  
dogs  dying  from  rabies  disease.

If $E_H\left(0\right)=0$, $I_H\left(0\right)=0$, $E_F\left(0\right)=0$, 
$I_F\left(0\right)=0$, $E_D\left(0\right)=1$, $I_D\left(0\right)=0$, 
and $M\left(0\right)=0$, then the offspring probability of generating  
function for  $E_D$ is given by
\begin{equation}
f_5\left(u_1, u_2, u_3, u_4,\;\ldots u_7,\;\right)
=\dfrac{\gamma_1 u_6+\gamma_2+\mu_3}{\gamma_1+\gamma_2+\mu_3},
\label{pgf5}
\end{equation} 
where  ${\gamma_1}\slash{\gamma_1+\gamma_2+\mu_3}$ represents  
the probability  of exposed domestic dog progressing  to infected  
class, ${\gamma_2}\slash{\gamma_1+\gamma_2+\mu_3}$  denotes the 
probability of exposed  domestic dogs  recovering  from exposed  
class  as a result of Rabies Postexposure Prophylaxis (PEP) 
before progressing to infected  class, and ${\mu_3}\slash{\gamma_1+\gamma_2+\mu_3}$  
signifies the probability of infected domestic dogs  
dying naturally before  progressing to infected  class.

The offspring probability generating function for $I_D$, such that 
$E_H\left(0\right)=0$, $I_H\left(0\right)=0$, $E_F\left(0\right)=0$, 
$I_F\left(0\right)=0$, $E_D\left(0\right)=0$, $I_D\left(0\right)=1$, 
and $M\left(0\right)=0$, is  given by
\begin{equation}
\begin{aligned}
f_6\left(u_1, u_2, u_3, u_4, \ldots, u_7\right)
=\dfrac{\hat{\lambda}_{4} u_5 u_6+\hat{\lambda}_{5} u_1 u_6
+\hat{\lambda}_{6} u_3 u_6+\mu_3+\sigma_3+\nu_3 u_6 u_7}{\hat{\lambda}_{4}
+\hat{\lambda}_{5}+\hat{\lambda}_{6}+\sigma_3+\mu_3+\nu_3},\;\; \\  
\textup{for}\; \hat{\lambda}_{4}=\dfrac{\psi_2}{1+\rho_2} S^{0}_{D},\;
\hat{\lambda}_{5}=\tau_2 S^{0}_{H},\;\hat{\lambda}_{6}=\kappa_2 S^{0}_{F}.
\end{aligned}
\label{pgf6}
\end{equation}
In pgf \eqref{pgf6}, the term  ${\hat{\lambda}_{4}}\slash{\hat{\lambda}_{4}
+\hat{\lambda}_{5}+\hat{\lambda}_{6}+\sigma_3+\mu_3+\nu_3}$ represents  
the probability  of infected domestic  dogs  causing  rabies infection  
to susceptible  domestic dogs, ${\hat{\lambda}_{5}}\slash{\hat{\lambda}_{4}
+\hat{\lambda}_{5}+\hat{\lambda}_{6}+\sigma_3+\mu_3+\nu_3}$ signifies 
the probability of a domestic dogs  causing infection to  susceptible humans, 
${\hat{\lambda}_{6}}\slash{\hat{\lambda}_{4}+\hat{\lambda}_{5}+\hat{\lambda}_{6}
+\sigma_3+\mu_3+\nu_3}$ denotes domestic dogs  causing infection 
to susceptible free range dogs, ${\nu_3}\slash{\hat{\lambda}_{4}
+\hat{\lambda}_{5}+\hat{\lambda}_{6}+\sigma_3+\mu_3+\nu_3}$ represents  
the probability of domestic dogs to shade rabies virus in the environment,  
${\mu_3}\slash{\hat{\lambda}_{4}+\hat{\lambda}_{5}+\hat{\lambda}_{6}
+\sigma_3+\mu_3+\nu_3}$ denotes the probability of infected domestic dogs  
dying naturally, and the term ${\sigma_3}\slash{\hat{\lambda}_{4}
+\hat{\lambda}_{5}+\hat{\lambda}_{6}+\sigma_3+\mu_3+\nu_3}$  is the 
probability of infected domestic  dogs  dying  from  rabies  disease.

If $E_H\left(0\right)=0$, $I_H\left(0\right)=0$, $E_F\left(0\right)=0$, 
$I_F\left(0\right)=0$, $E_D\left(0\right)=0$, $I_D\left(0\right)=0$, 
and $M\left(0\right)=1$, then the offspring probability of generating  
function for  $M$ is given by
\begin{equation}
\begin{aligned}
f_7\left(u_1\;,u_2\;,u_3\;,u_4,\;\ldots, u_7\;\right)
=\dfrac{\hat{\lambda}_{7} u_5 u_7+\hat{\lambda}_{8} u_1 u_7
+\hat{\lambda}_{9} u_3 u_7+\mu_4}{\hat{\lambda}_{7}
+\hat{\lambda}_{8}+\hat{\lambda}_{9}+\mu_4},\;\; \\  
\textup{for}\; \hat{\lambda}_{7}=\dfrac{\psi_3}{1+\rho_3} 
S^{0}_{D},\;\hat{\lambda}_{8}=\tau_3 S^{0}_{H},\;\hat{\lambda}_{9}
=\kappa_3 S^{0}_{F},
\end{aligned}
\label{pgf7}
\end{equation}
where ${\hat{\lambda}_{7}}\slash{\hat{\lambda}_{7}+\hat{\lambda}_{8}
+\hat{\lambda}_{9}+\mu_4}$ represents  the probability  of rabies 
in environment causing infection  to susceptible  domestic dogs, 
${\hat{\lambda}_{8}}\slash{\hat{\lambda}_{7}+\hat{\lambda}_{8}
+\hat{\lambda}_{9}+\mu_4}$ signifies the probability of rabies 
in environment causing infection  to  susceptible  humans, 
${\hat{\lambda}_{9}}\slash{\hat{\lambda}_{7}+\hat{\lambda}_{8}
+\hat{\lambda}_{9}+\mu_4}$ denotes the probability of rabies 
in environment causing infection to susceptible free range dogs,  
and ${\mu_4}\slash{\hat{\lambda}_{7}+\hat{\lambda}_{8}+\hat{\lambda}_{9}
+\mu_4}$ denotes the probability of removal of  rabies in environment.  

The expectation matrix \( \mathbb{M} \) of the branching process is a 
\( 7 \times 7 \) matrix, which is defined by equation \eqref{M1}. 
It is derived from the offspring probability generating functions (pgfs) 
given in equations \eqref{pgf1} to \eqref{pgf7}, with all variables 
\((u_1, u_2, u_3, u_4, u_5, u_6, u_7)=(1, 1, 1, 1, 1, 1, 1) \):
\begin{equation}
\mathbb{M} = 
\begin{pmatrix}
0 & 0 & 0 & \dfrac{\hat{\lambda}_{1}}{J_1} 
& 0& \dfrac{\hat{\lambda}_{5}}{J_2}& \dfrac{\hat{\lambda}_{8}}{J_3}\\\\
\dfrac{\beta_1}{J_4} & \dfrac{\nu_1}{J_5} & 0 & 0 & 0 & 0& 0 \\\\
0 & 0 & 0 & \dfrac{\hat{\lambda}_{3}}{J_6} &0 & \dfrac{\hat{\lambda}_{6}}{J_7}
& \dfrac{\hat{\lambda}_{9}}{J_8} \\\\
0 & 0 & \dfrac{\gamma}{\left(\gamma+\mu_{2}\right)} 
& \dfrac{\mathbb{G}_1}{J_1} &0 & 0& 0 \\\\
0 & 0 & 0 & \dfrac{\hat{\lambda}_{4}}{J_1}&0& \dfrac{\hat{\lambda}_{4}}{J_2}
& \dfrac{\hat{\lambda}_{7}}{J_3} \\\\
0 & 0 & 0 & 0 & \dfrac{\gamma_{1}}{\left(\gamma_{1}+\gamma_{2}+\mu_{3}\right)} 
& \dfrac{\mathbb{G}_2}{J_2}& 0 \\\\
0 & \dfrac{\nu_{1}}{\left(\nu_{1}+\mu_{1}+\sigma_{1}\right)} 
& 0& \dfrac{\nu_{2}}{J_{1}} & 0 
& \dfrac{\nu_{3}}{J_2}& \dfrac{\hat{\lambda}_{9}}{J_3} \\
\end{pmatrix},
\label{M1}
\end{equation}
where
\begin{gather*}
J_1=\hat{\lambda}_1+\hat{\lambda}_2+\hat{\lambda}_3
+\sigma_{2}+\mu_{2}+\nu_2,\; J_2=\hat{\lambda}_4
+\hat{\lambda}_5+\hat{\lambda}_6+\sigma_{3}+\mu_{3}+\nu_3,\; 
J_3=\hat{\lambda}_7+\hat{\lambda}_8+\hat{\lambda}_9+\mu_{4},\\
J_4=\beta_1+\beta_2+\mu_1,\; J_5=\nu_1+\sigma_1+\mu_1,\; 
J_6=\hat{\lambda}_1+\hat{\lambda}_2+\hat{\lambda}_3+\mu_{2}+\sigma_{2}+\nu_{2},\\  
J_7=\hat{\lambda}_4+\hat{\lambda}_5+\hat{\lambda}_6+\mu_{3}+\sigma_{3}+\nu_{3},\; 
J_8=\hat{\lambda}_7+\hat{\lambda}_8+\hat{\lambda}_9+\mu_{4},\; 
\mathbb{G}_1=\hat{\lambda}_{1}+\hat{\lambda}_{2}+\hat{\lambda}_{3}
+\nu_2,\; \mathbb{G}_2=\hat{\lambda}_{4}+\hat{\lambda}_{5}+\hat{\lambda}_{6}+\nu_3.
\end{gather*}
The Continuous-Time Markov Chain (CTMC) model identifies a stochastic 
threshold that determines whether rabies will die out or lead to an outbreak 
in human and dog populations. This threshold is represented by the spectral 
radius of the expectation matrix, \(\rho\left(\mathbb{M}\right)\). There 
is a relationship between \(\rho\left(\mathbb{M}\right)\) in the stochastic 
model and the basic reproduction number \(\mathcal{R}_{0}\) in the deterministic 
model. For rabies to be eliminated from both human and dog populations, 
it is required that \(\rho\left(\mathbb{M}\right)\leq 1\) or \(\mathcal{R}_{0}<1\). 
Conversely, in deterministic models, rabies persists in humans and dogs if 
\(\mathcal{R}_{0} > 1\). The relationship between the deterministic and 
stochastic thresholds for rabies extinction can be expressed as 
\(\mathcal{R}_{0} < 1 \iff \rho\left(\mathbb{M}\right) < 1\).  
In stochastic models, when \(\rho\left(\mathbb{M}\right) < 1\), there 
is a possibility of either an outbreak or extinction of the 
\textit{Rabies lyssavirus}, depending on the initial number 
of infectives at the onset of the disease outbreak. Conversely, 
if \(\rho\left(\mathbb{M}\right) > 1\), a fixed point 
\[
\left(f_1,\;f_2,\;f_3,\;f_4,\;f_5,\;f_6,\;f_7\right) \in \left(0,\;1\right)^6
\]
can be determined using offspring generating functions, which are then 
used to assess the probability of disease extinction. These generating 
functions are nonlinear, making analytical computation challenging, 
and thus numerical methods are typically employed for their calculation.


\section{Quantitative Analysis: Numerical Simulations}
\label{sec:03}

Following the analytical assessment of both the deterministic and Continuous-Time Markov Chain (CTMC) 
stochastic frameworks, numerical simulations were conducted to investigate the qualitative dynamics 
of the proposed rabies transmission model. Accurate parameter estimation is essential for generating 
reliable quantitative forecasts within constrained time frames using real-world epidemiological data 
\citep{charles2024mathematical}.  Model parameters in equation~\eqref{eq:stochastic_model_with_noise} 
were estimated using the non-linear least squares method. Synthetic datasets were obtained 
by numerically integrating equation~\eqref{eq:stochastic_model_with_noise} with a fifth-order 
Runge--Kutta scheme implemented in \textsf{MATLAB}, employing initial parameter values 
$\Theta_i$ from the literature and the initial population conditions:
\[
\begin{aligned}
& S_H(0) = 142{,}000,\quad E_H(0) = 40,\quad I_H(0) = 0,\quad R_H(0) = 0, \\
& S_D(0) = 15{,}000,\quad E_D(0) = 25,\quad I_D(0) = 0,\quad R_D(0) = 0, \\
& S_F(0) = 12{,}500,\quad E_F(0) = 20,\quad I_F(0) = 0,\quad M(0) = 90.
\end{aligned}
\]

The observed data were formalized as a stochastic process:
\[
Y_i = RD(t_i,\Theta_i) + \eta_i,\quad \eta_i 
\overset{\text{i.i.d.}}{\sim} \mathcal{N}(0,\sigma^2), 
\quad t_i \in [1,n],
\]
where \( RD(t_i,\Theta_i) \) denotes the model-predicted prevalence 
and \( \eta_i \) represents Gaussian measurement noise. Parameter 
estimates were obtained under the assumption that deviations from 
baseline literature values follow a Gaussian distribution, 
as reported in Table~\ref{T2}.

\begin{center}
\begin{longtable}{|l|l|l|l|l|}
\caption{\centering Estimated model parameters (Year$^{-1}$), 
initial guess for parameters (Year$^{-1}$) and their respective source.}\\  \hline 
\textbf{Parameters} & \textbf{Baseline value} & \textbf{Source} 
& \textbf{Estimated value}& $\textup{Mean}\left(\mu \right) 
\textup{and std}\left(\sigma\right) $\\ \hline\hline
\endfirsthead
\multicolumn{4}{c}
{\tablename\ \thetable\ -- \textit{Continued from previous page}} \\ \hline
\textbf{Parameters} & \textbf{Baseline value} & \textbf{Source} 
& \textbf{Estimated value}& $\textup{Mean}\left(\mu \right) 
\textup{and std}\left(\sigma\right) $\\ \hline
\endhead
\hline \multicolumn{4}{r}{\textit{Continued on next page}} \\
\endfoot
\hline
\endlastfoot
$\theta_{1}$ & 2000& (Assumed)&1993.382113 
& $\mathcal{ N}\left(1996.691056\;\; 4.4679553 \right)$ \\
$\tau_{1}$ & 0.0004& \cite{tian2018transmission}&0.000405 
&$\mathcal{ N}\left(0.000402\;\; 4\times10^{-6} \right)$\\
$\tau_{2}$ & 0.0004& \cite{tian2018transmission}&0.000604
&$\mathcal{ N}\left(0.000502\;\; 1.44\times10^{-4} \right)$\\
$\tau_{3}$ & $\left[0.0003\;\;  0.0100\right]$ & (Assumed)
&0.000303 &$\mathcal{ N}\left(0.000302\;\; 2\times10^{-6} \right)$\\
$\beta_{1}$ & $\frac{1}{6}$ & \cite{tian2018transmission,zhang2011analysis}
&0.165581& $\mathcal{ N}\left(0.166124\;\; 7.68\times10^{-4} \right)$\\
$\beta_{2}$ & $\left[0.54 \;\; 1\right]$ & \cite{zhang2011analysis,abdulmajid2021analysis}
&0.540487&$\mathcal{ N}\left(0.5402435\;\; 3.7815\times10^{-4} \right)$\\
$\beta_{3}$ & 1&(Assumed)&0.999301&$\mathcal{ N}\left(0.9996505\;\; 1.6521\times10^{-4} \right)$\\
$\mu_1$ & 0.0142& \cite{world2013expert}&0.014417&$\mathcal{ N}\left(0.014309\;\; 1.53\times10^{-4} \right)$\\
$\sigma_1$ & 1& \cite{zhang2011analysis,abdulmajid2021analysis} &1.006332
&$\mathcal{ N}\left(1.03166\;\; 4.47\times10^{-3} \right)$\\
$\theta_{2}$ & 1000& (Assumed)&1004.12044&$\left(1002.060222\;\; 2.913594\right)$\\
$\kappa_{1}$ &  0.00006 & (Assumed)&0.000020&$\mathcal{ N}\left(0.000040\;\; 2.8\times10^{-5} \right)$\\
$\kappa_{2}$ & 0.00005 & (Assumed)&0.000081&$\mathcal{ N}\left(0.000066\;\; 2.2\times10^{-5} \right)$\\
$\kappa_{3}$ & $\left[0.00001 \;\; 0.00003\right]$&(Assumed)
&0.000040&$\mathcal{ N}\left(0.000025\;\; 2.1\times10^{-5} \right)$\\
$\gamma$ & $\frac{1}{6}$ & \cite{tian2018transmission,zhang2011analysis,abdulmajid2021analysis} 
& 0.166374&$\mathcal{ N}\left(0.166520\;\; 2.07\times10^{-4} \right)$\\
$\sigma_2$ & 0.09 & \cite{zhang2011analysis,addo2012seir}
&0.089556&$\mathcal{ N}\left(0.089778\;\; 3.14\times10^{-4} \right)$\\
$\mu_{2}$ & 0.067 & (Assumed)&0.066268 
&$\mathcal{ N}\left(0.066634\;\; 1.58\times10^{-4} \right)$\\
$\theta_3$ & 1200& (Assumed)&1203.844461&$\mathcal{ N}\left(1201.922230\;\; 2.718444 \right)$\\
$\psi_{1}$ &0.0004& \cite{addo2012seir,hampson2019potential}
&0.000077&$\mathcal{ N}\left(0.000238\;\; 2.28\times10^{-4} \right)$\\
$\psi_{2}$ & 0.0004 & \cite{CHARLES2024115633} &0.000066
&$\mathcal{ N}\left(0.000233\;\; 2.36\times10^{-4} \right)$\\
$\psi_{3}$ & 0.0003 & (Assumed)&0.000030&$\mathcal{ N}\left(0.0003\;\; 1.91\times10^{-4} \right)$\\
$\mu_3$ &0.067& (Assumed)&0.080129&$\mathcal{ N}\left(0.073565\;\; 8.056\times10^{-3} \right)$\\
$\sigma_3$ &0.08& \cite{zhang2011analysis}&0.091393 & $\mathcal{ N}\left(0.085697\;\; 8.056\times10^{-3} \right)$\\
$\gamma_1$ & $\frac{1}{6}$ & \cite{tian2018transmission,zhang2011analysis}
&0.172489&$\mathcal{ N}\left(0.169578\;\; 4.117\times10^{-3} \right)$\\
$\gamma_2$ & 0.09 & \cite{zhang2011analysis}&0.090308
&$\mathcal{ N}\left(0.090154\;\; 2.18\times10^{-4} \right)$\\
$\gamma_3$ & 0.05&(Assumed)&0.050128&$\mathcal{ N}\left(0.050128\;\; 9.1\times10^{-5} \right)$\\
$\nu_1$ &0.001& (Assumed)&0.001958&$\mathcal{ N}\left(0.001479\;\; 6.77\times10^{-4} \right)$\\
$\nu_2$ &0.006&(Assumed)&0.008971&$\mathcal{ N}\left(0.007485\;\; 2.101\times10^{-3} \right)$\\
$\nu_3$ &0.001& (Assumed)&0.005735&$\mathcal{ N}\left(0.003367\;\; 3.3348\times10^{-3} \right)$\\
$\mu_4$  &0.08& (Assumed)&0.080625&$\mathcal{ N}\left(0.080313\;\; 4.42\times10^{-4} \right)$\\
$\rho_{1}$ & 10& \cite{charles2024mathematical}&9.920733
&$\mathcal{ N}\left(9.960366\;\; 5.605\times10^{-2} \right)$\\
$\rho_{2}$ & 8& (Assumed)&8.116421&$\mathcal{ N}\left(8.058211\;\; 8.2322\times10^{-2} \right)$\\
$\rho_{3}$ & 15 &(Assumed)&14.917005 &$\mathcal{ N}\left(14.958502\;\; 5.8686\times10^{-2} \right)$\\
$C$  &0.003  (PFU)/mL & (Assumed)&0.003011&$\mathcal{ N}\left(0.003005\;\; 8.0000\times10^{-6} \right)$
\label{T2}
\end{longtable}
\end{center}

The simulation is performed using 10,000 random sample paths, with the results 
presented graphically alongside the corresponding deterministic numerical 
solutions for comparative analysis. To conduct the simulations, Euler's 
method and the Gillespie algorithm are utilized, applying the specified 
initial conditions \(E_H\left(0\right)=10\), \(I_H\left(0\right)=5\),
\(R_H\left(0\right)=0\), \(S_H\left(0\right)=\dfrac{\theta_{1}}{\mu_{1}}
-\left(E_H\left(0\right)+I_H\left(0\right)+R_H\left(0\right)\right)\), 
\(E_F\left(0\right)=20\), \(I_F\left(0\right)=0\),
\(S_F\left(0\right)=\dfrac{\theta_{2}}{\mu_{2}}
-\left(E_F\left(0\right)+I_F\left(0\right)\right)\), 
\(E_D\left(0\right)=40\), \(I_H\left(0\right)=5\),\(R_D\left(0\right)=0\),
and \(S_D\left(0\right)=\dfrac{\theta_{3}}{\mu_{3}}-\left(E_D\left(0\right)
+I_D\left(0\right)+R_D\left(0\right)\right)\).

\begin{figure}[H]
\begin{minipage}[b]{0.45\textwidth}
\includegraphics[height=5.0cm, width=7.5cm]{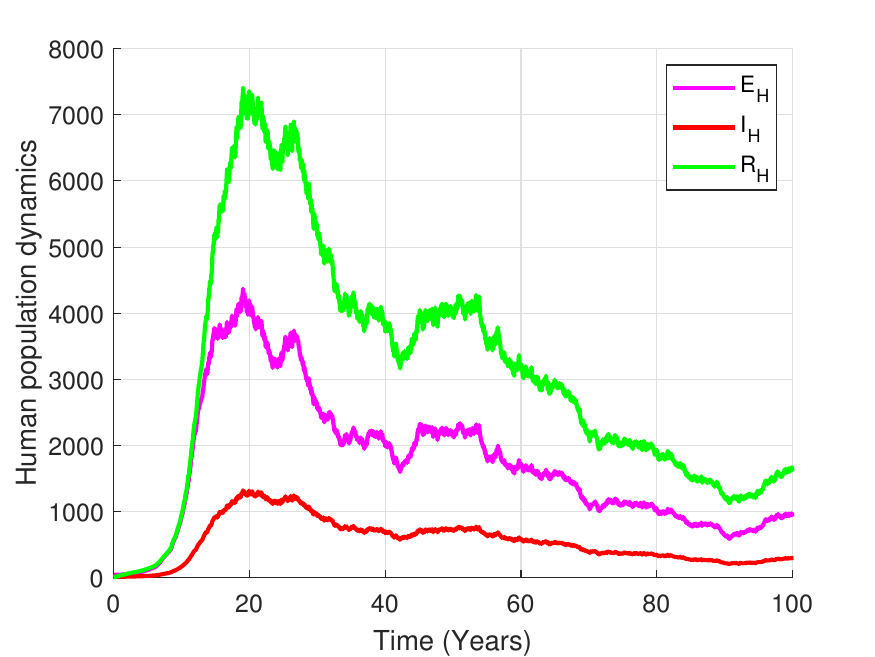}
\centering{(a)}
\end{minipage}
\begin{minipage}[b]{0.45\textwidth}
\includegraphics[height=5.0cm, width=7.5cm]{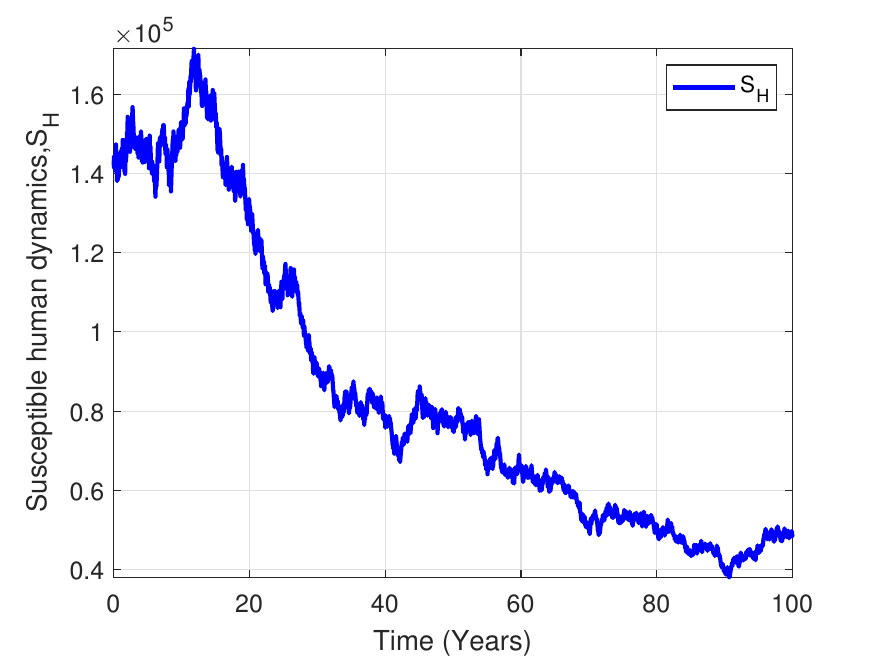}
\centering{(b)}
\end{minipage}
\centering  
\caption{\centering Stochastic rabies transmission dynamics in humans.}
\label{Fig4}
\end{figure}

\begin{figure}[H]
\begin{minipage}[b]{0.45\textwidth}
\includegraphics[height=5.0cm, width=7.5cm]{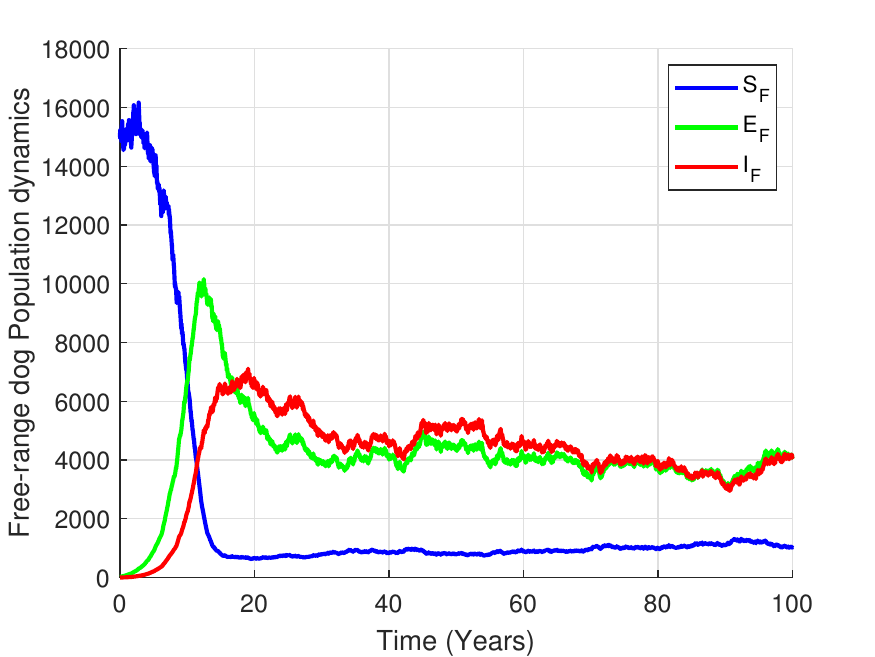}
\centering{(a)}
\end{minipage}
\begin{minipage}[b]{0.45\textwidth}
\includegraphics[height=5.0cm, width=7.5cm]{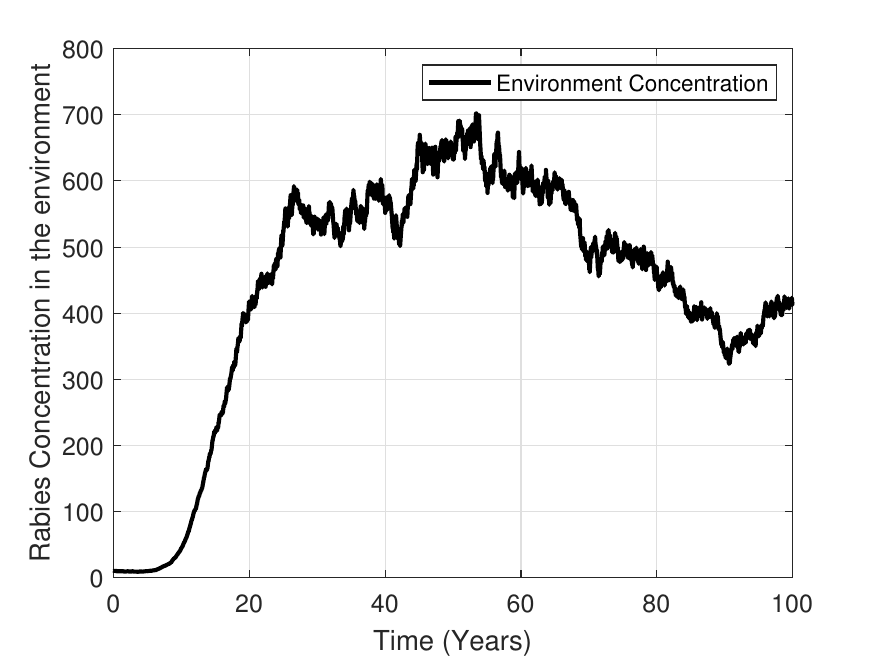}
\centering{(b)}
\end{minipage}
\centering  
\caption{\centering Stochastic rabies transmission dynamics 
in (a) Free range dogs (b) Environment.}
\label{Fig5}
\end{figure}

\begin{figure}[H]
\begin{minipage}[b]{0.45\textwidth}
\includegraphics[height=5.0cm, width=7.5cm]{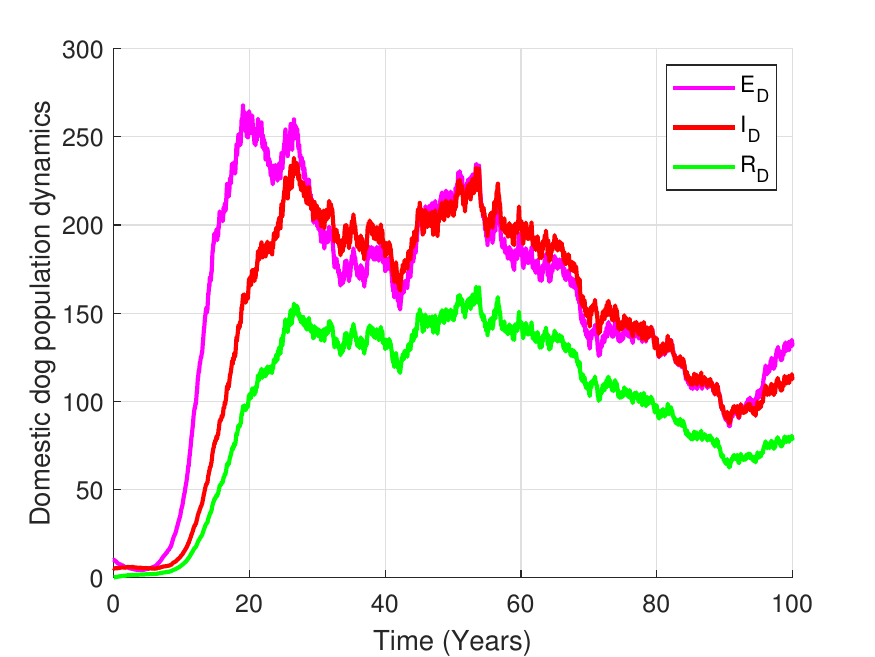}
\centering{(a)}
\end{minipage}
\begin{minipage}[b]{0.45\textwidth}
\includegraphics[height=5.0cm, width=7.5cm]{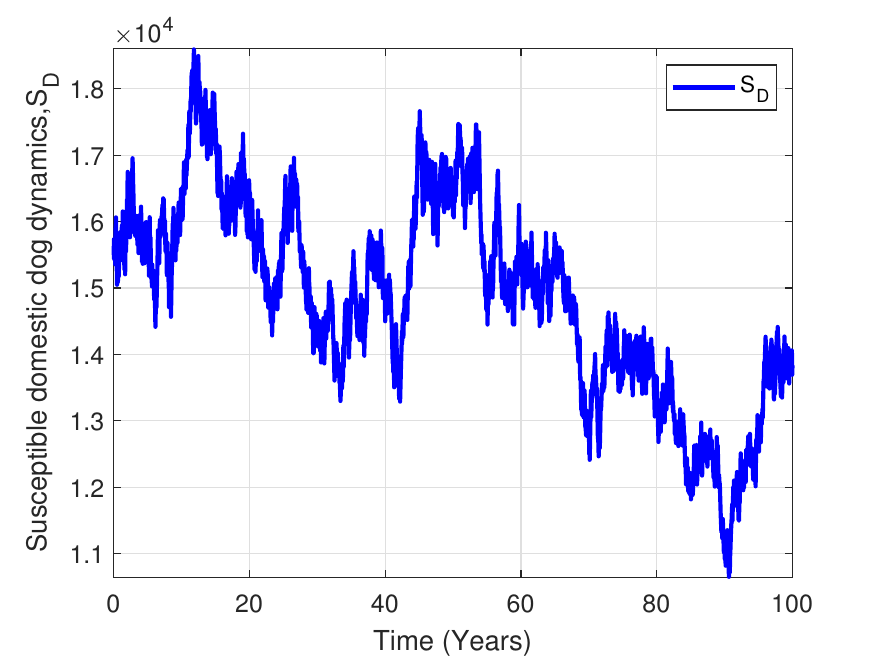}
\centering{(b)}
\end{minipage}
\centering  
\caption{\centering Stochastic rabies transmission dynamics in Domestic dogs.}
\label{Fig6}
\end{figure}

Figures~\ref{Fig4}--\ref{Fig6} illustrate the stochastic 
transmission dynamics of rabies between human and dog populations. 

Figure~\ref{Fig4}(b) illustrates that the number of susceptible humans initially 
experiences a stochastic decline over the first 40 years, eventually stabilizing 
at a variable, non-constant level. In contrast, Figure~\ref{Fig4}(a) depicts an 
increase in stochastic fluctuations among the exposed, infected, and recovered 
human populations during the same period, followed by a decline that approaches 
a steady, non-zero value. The observed fluctuations in the exposed, infected, 
and recovered populations likely reflect the effects of control interventions, 
such as the administration of post-exposure prophylaxis (PEP) to individuals 
exposed to rabid animals. Figure~\ref{Fig5}(a) shows that the number of susceptible 
free-range dogs initially decreases as the populations of exposed and infected 
free-range dogs rise stochastically over the first 20 years, eventually stabilizing 
at a variable, non-constant level. At the same time, Figure~\ref{Fig5}(b) indicates 
increasing stochastic fluctuations in environmental rabies concentration, 
which subsequently stabilize toward a steady, non-zero level. 

Finally, Figure~\ref{Fig6}(a) demonstrates that the number of susceptible 
domestic dogs initially undergoes periodic declines, while the population 
of exposed, infected, and recovered domestic dogs in Figure~\ref{Fig6}(b) 
increases stochastically over the first 20 years, eventually reaching 
stability at a variable, non-constant level.
\begin{figure}[H]
\begin{minipage}[b]{0.45\textwidth}
\includegraphics[height=5.0cm, width=7.5cm]{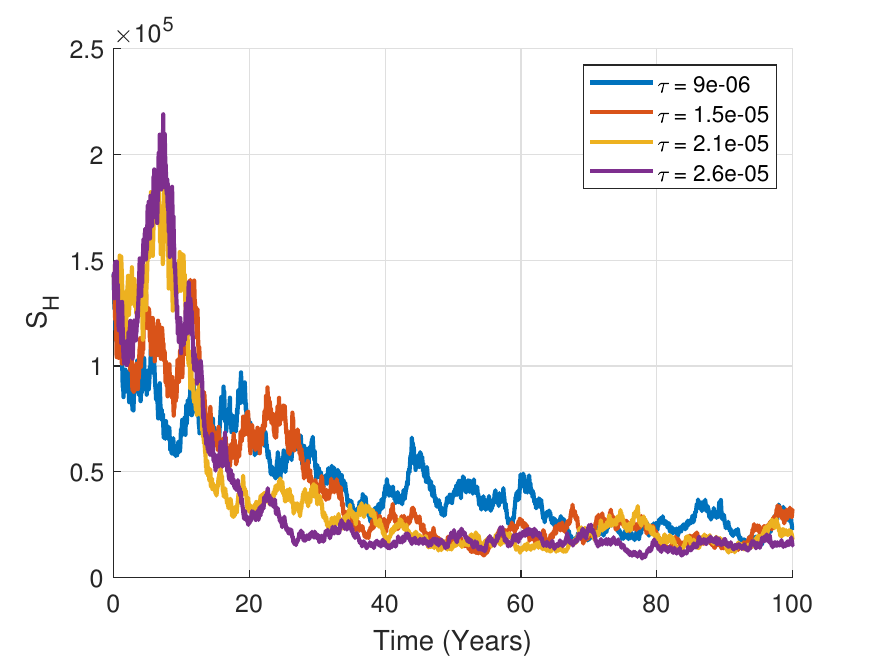}
\centering{(a)}
\end{minipage}
\begin{minipage}[b]{0.45\textwidth}
\includegraphics[height=5.0cm, width=7.5cm]{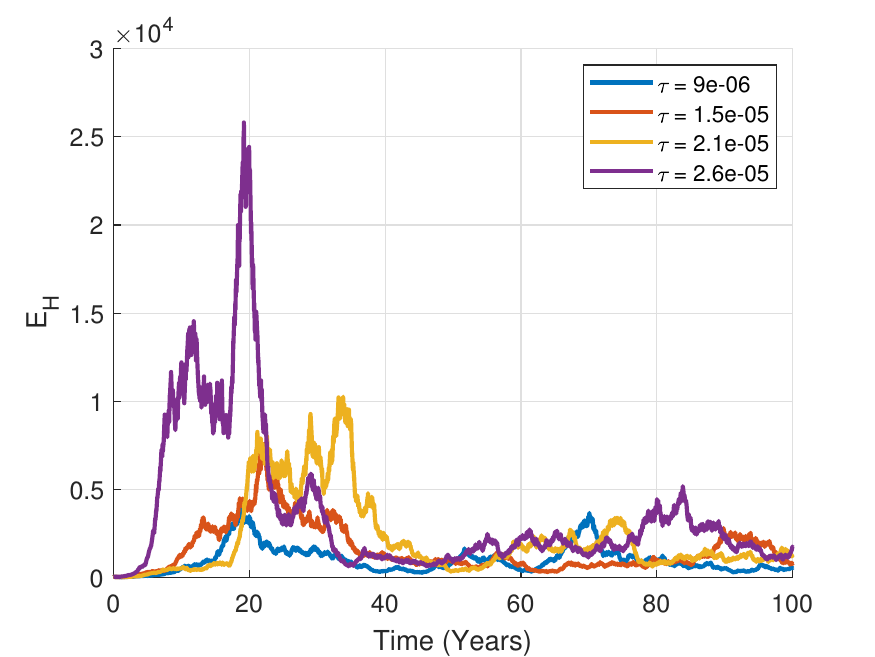}
\centering{(b)}
\end{minipage}
\begin{minipage}[b]{0.45\textwidth}
\includegraphics[height=5.0cm, width=7.5cm]{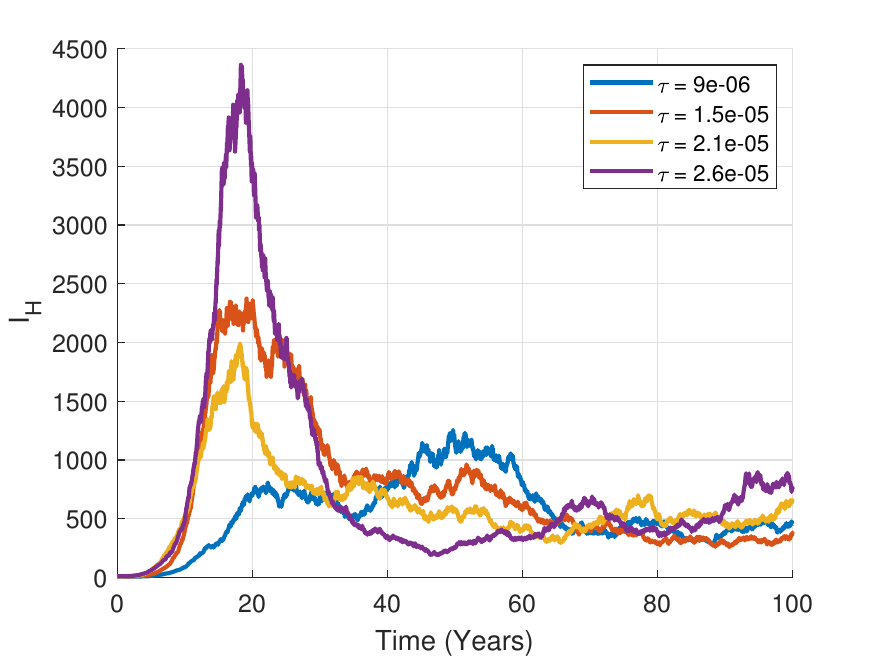}
\centering{(c)}
\end{minipage}
\centering
\caption{Stochastic trajectory of human  
due to impact of contact rate \(\tau\).}
\label{tau}
\end{figure}
\begin{figure}[H]
\begin{minipage}[b]{0.45\textwidth}
\includegraphics[height=4.0cm, width=6.5cm]{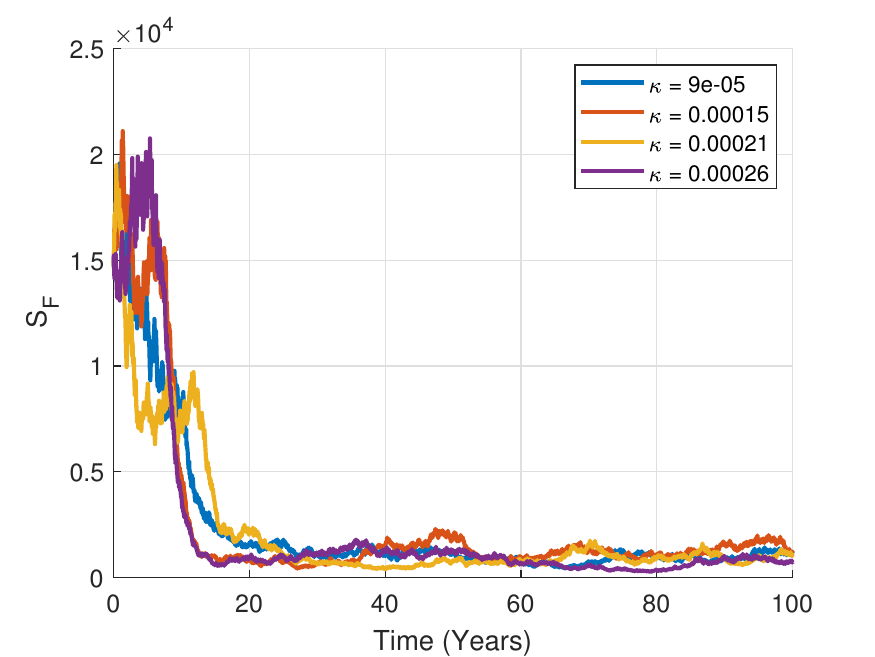}
\centering{(a)}
\end{minipage}
\begin{minipage}[b]{0.45\textwidth}
\includegraphics[height=4.0cm, width=6.5cm]{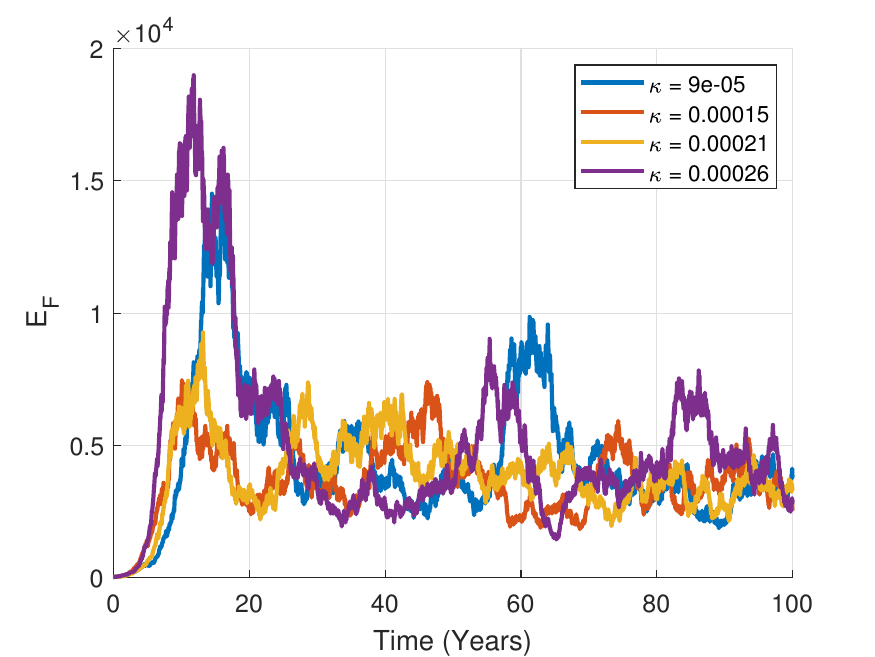}
\centering{(b)}
\end{minipage}
\begin{minipage}[b]{0.45\textwidth}
\includegraphics[height=4.0cm, width=6.5cm]{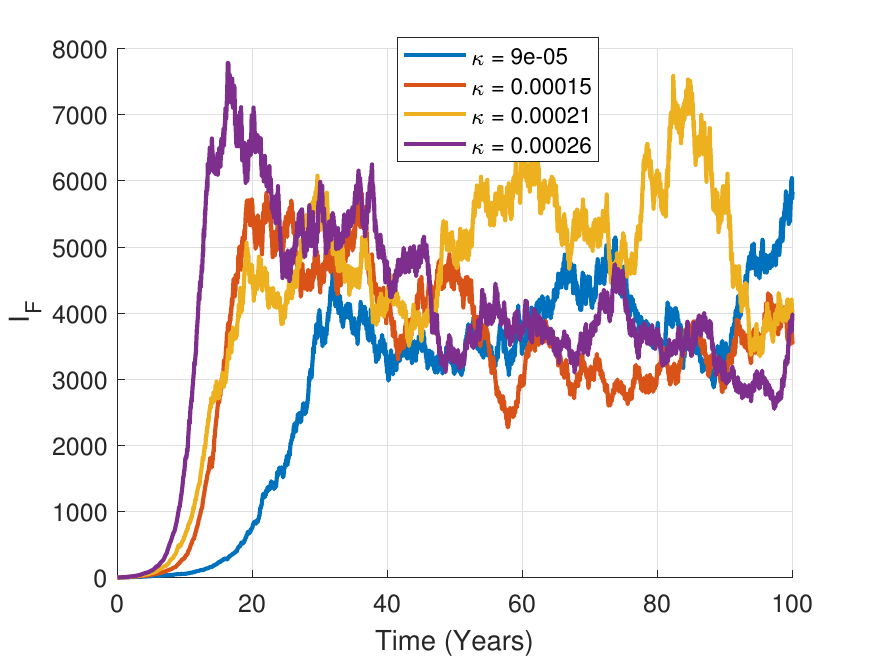}
\centering{(c)}
\end{minipage}
\centering
\caption{Stochastic trajectory of Free range dogs 
due to impact of contact rate \(\kappa\).}
\label{kappa}
\end{figure}
\begin{figure}[H]
\begin{minipage}[b]{0.45\textwidth}
\includegraphics[height=4.0cm, width=6.5cm]{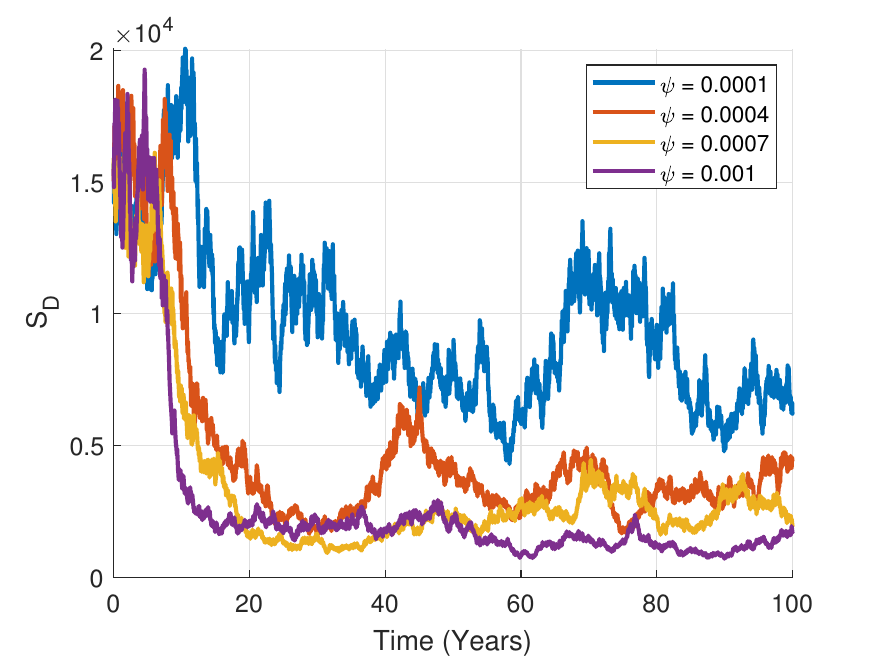}
\centering{(a)}
\end{minipage}
\begin{minipage}[b]{0.45\textwidth}
\includegraphics[height=4.0cm, width=6.5cm]{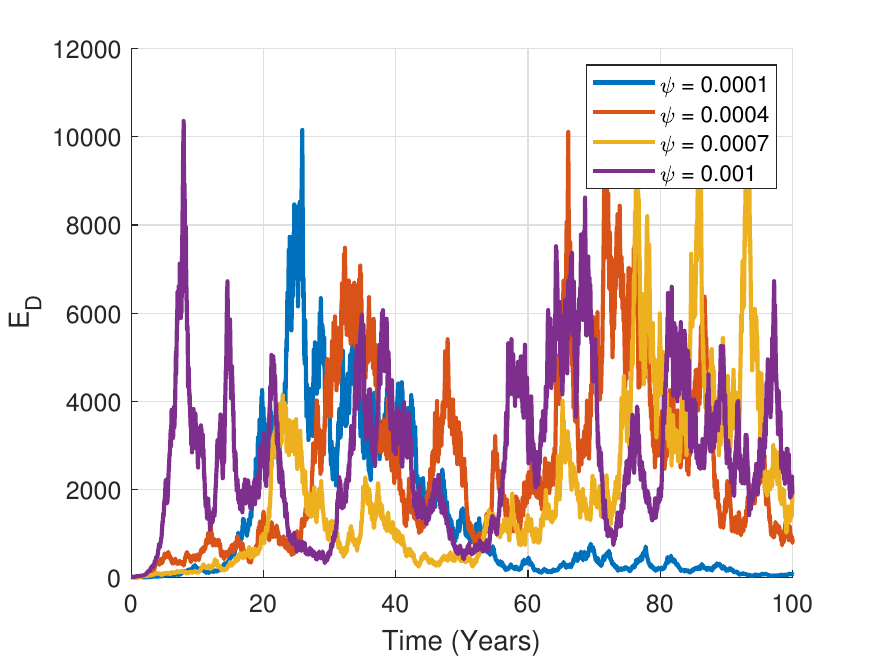}
\centering{(b)}
\end{minipage}
\begin{minipage}[b]{0.45\textwidth}
\includegraphics[height=4.0cm, width=6.5cm]{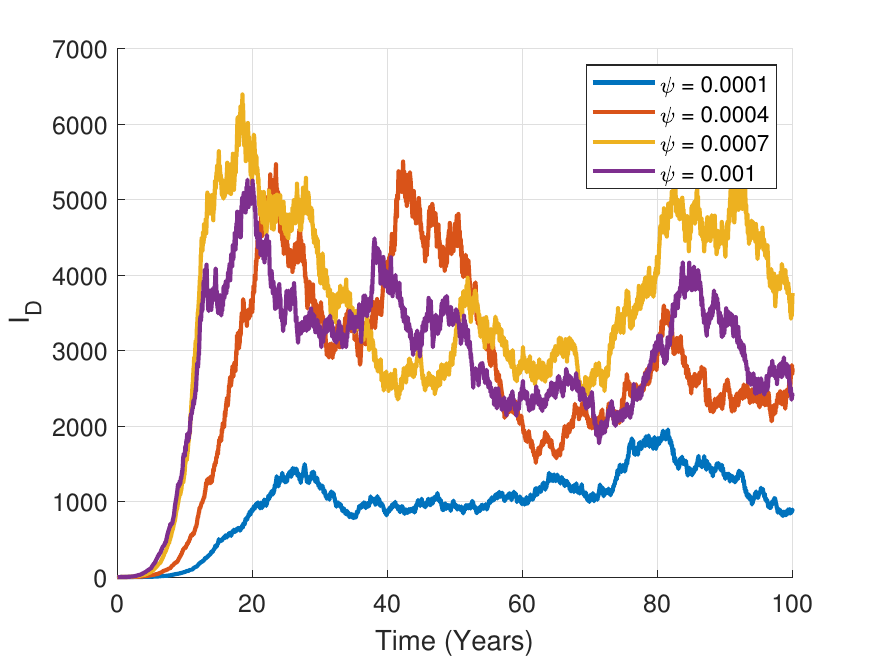}
\centering{(c)}
\end{minipage}
\centering
\caption{Stochastic trajectory of domestic dogs  
due to impact of contact rate  \(\psi\).}
\label{psi}
\end{figure}
\begin{figure}[H]
\centering
\includegraphics[scale=.6]{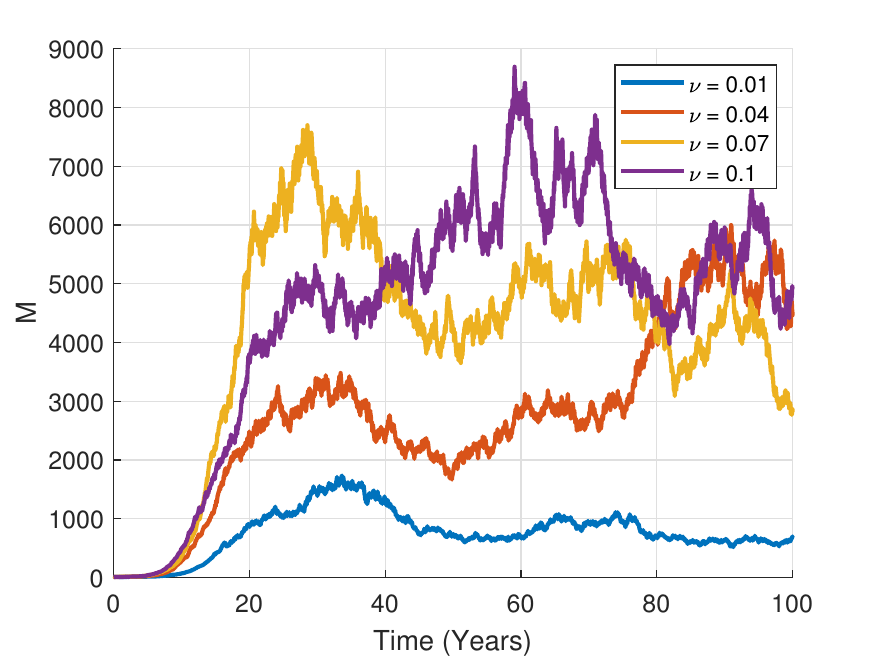}
\caption{Stochastic trajectory of Environment 
due to impact of shedding rate  \(\nu\).}
\label{nu}
\end{figure}

Figures~\eqref{tau} through \eqref{psi} clearly illustrate a stochastic rise in the number 
of infected individuals among humans, domestic animals, and free-range dogs, while 
concurrently, there is a decline in the number of susceptible individuals within 
these populations. Additionally, there is a simultaneous stochastic increase 
in the concentration of rabies in the environment, driven by variations 
in contact and shedding rates. These scenarios provide evidence that the movement 
of free-roaming dogs could potentially introduce a new rabies infection, 
suggesting a risk of an outbreak.

\begin{figure}[H]
\begin{minipage}[b]{0.45\textwidth}
\includegraphics[height=5.0cm, width=7.5cm]{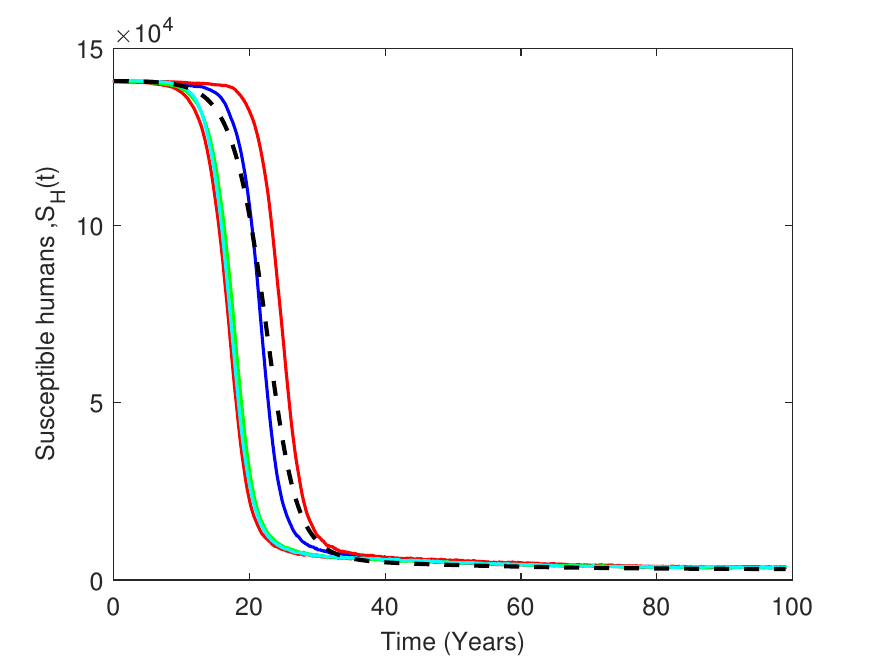}
\centering{(a)}
\end{minipage}
\begin{minipage}[b]{0.45\textwidth}
\includegraphics[height=5.0cm, width=7.5cm]{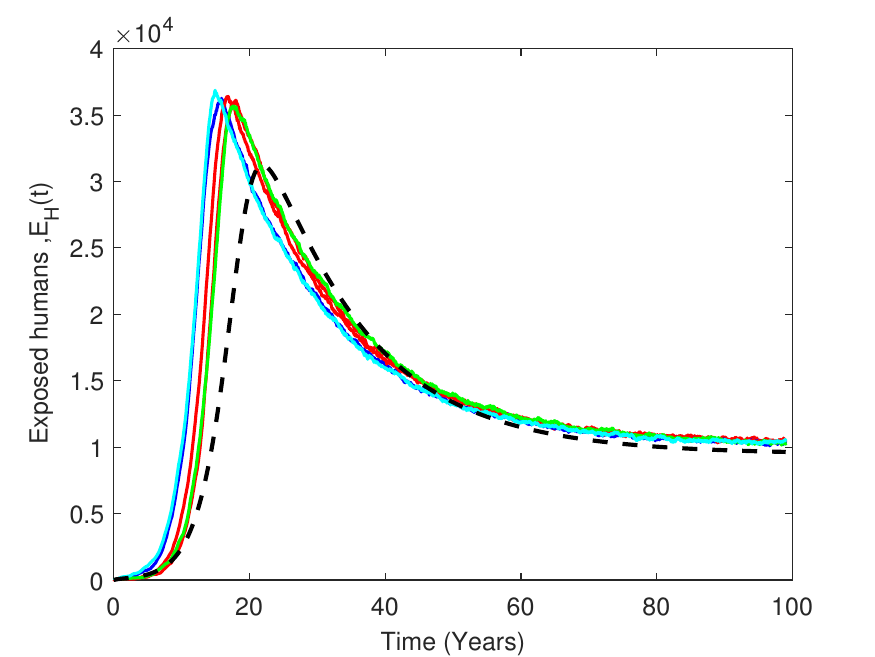}
\centering{(b)}
\end{minipage}
\centering  
\caption{\centering Comparison of Deterministic (dotted lines) 
and CTMC Sample Paths for Rabies Transmission Dynamics: 
(a)~Susceptible Humans and (b)~Exposed Humans.}
\label{Fig7}
\end{figure}

\begin{figure}[H]
\begin{minipage}[b]{0.45\textwidth}
\includegraphics[height=5.0cm, width=7.5cm]{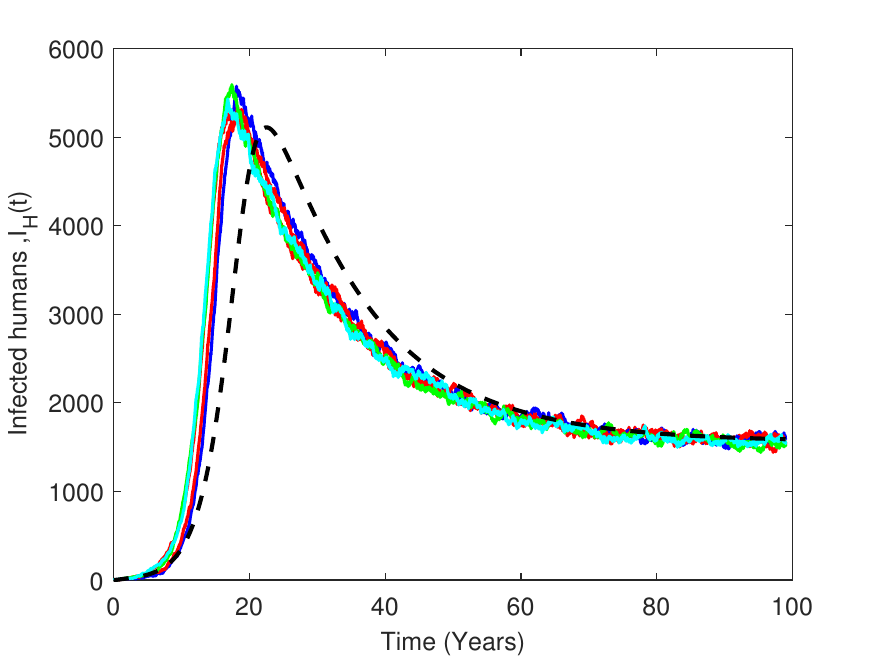}
\centering{(a)}
\end{minipage}
\begin{minipage}[b]{0.45\textwidth}
\includegraphics[height=5.0cm, width=7.5cm]{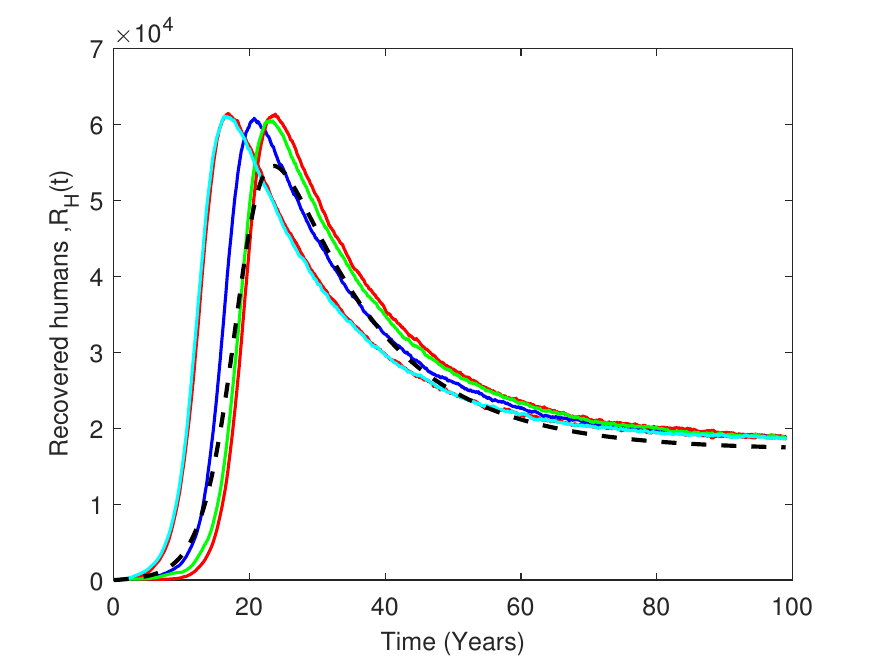}
\centering{(b)}
\end{minipage}
\centering  
\caption{\centering Comparison of Deterministic (dotted lines) 
and CTMC Sample Paths for Rabies Transmission Dynamics: 
(a)~Infected Humans and (b)~Recovered Humans.}
\label{Fig8}
\end{figure}

\begin{figure}[H]
\begin{minipage}[b]{0.45\textwidth}
\includegraphics[height=5.0cm, width=7.5cm]{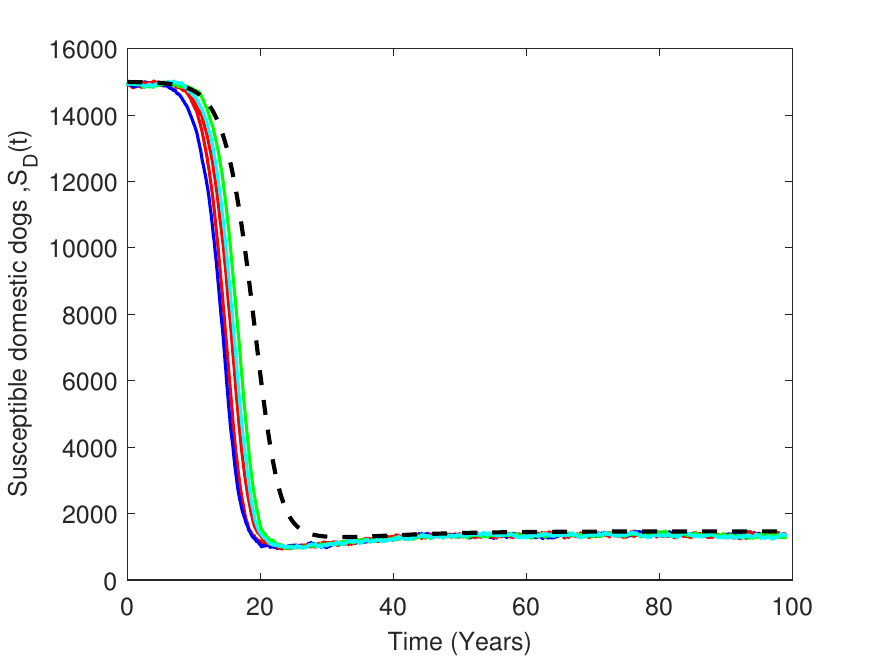}
\centering{(a)}
\end{minipage}
\begin{minipage}[b]{0.45\textwidth}
\includegraphics[height=5.0cm, width=7.5cm]{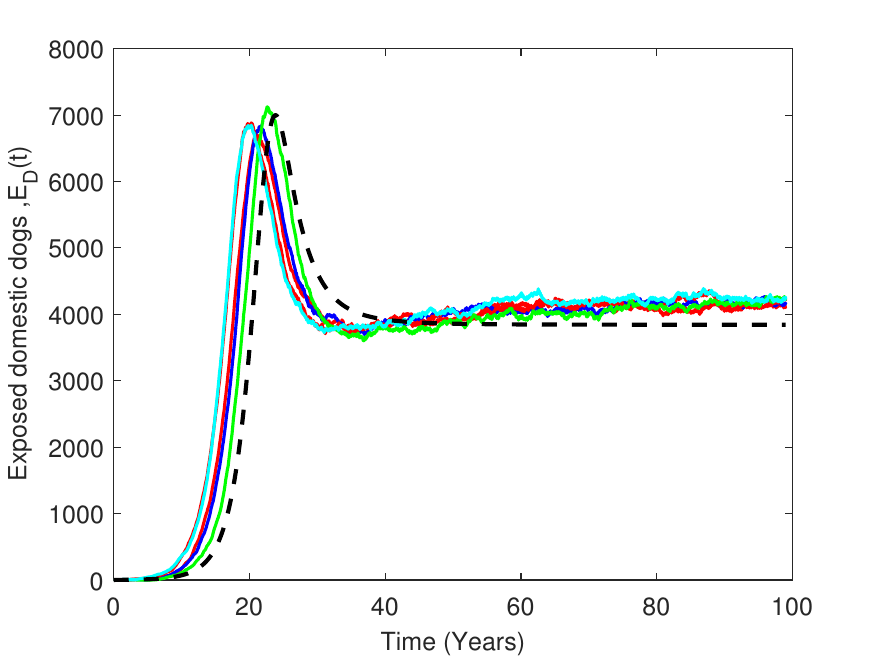}
\centering{(b)}
\end{minipage}
\centering  
\caption{\centering Comparison of Deterministic (dotted lines) 
and CTMC Sample Paths for Rabies Transmission Dynamics: 
(a)~Susceptible domestic dogs and (b)~Exposed domestic dogs.}
\label{Fig9}
\end{figure}

\begin{figure}[H]
\begin{minipage}[b]{0.45\textwidth}
\includegraphics[height=5.0cm, width=7.5cm]{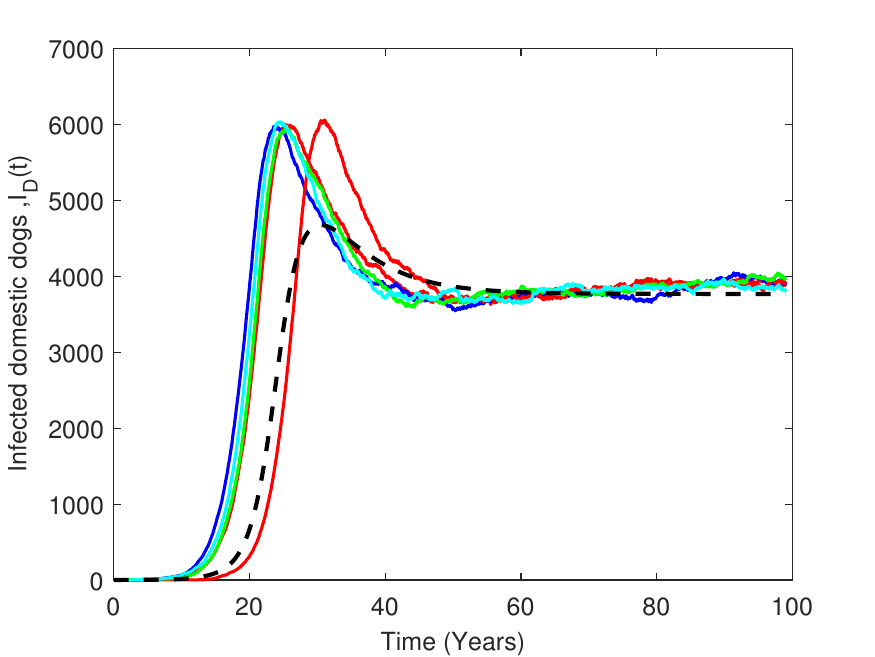}
\centering{(a)}
\end{minipage}
\begin{minipage}[b]{0.45\textwidth}
\includegraphics[height=5.0cm, width=7.5cm]{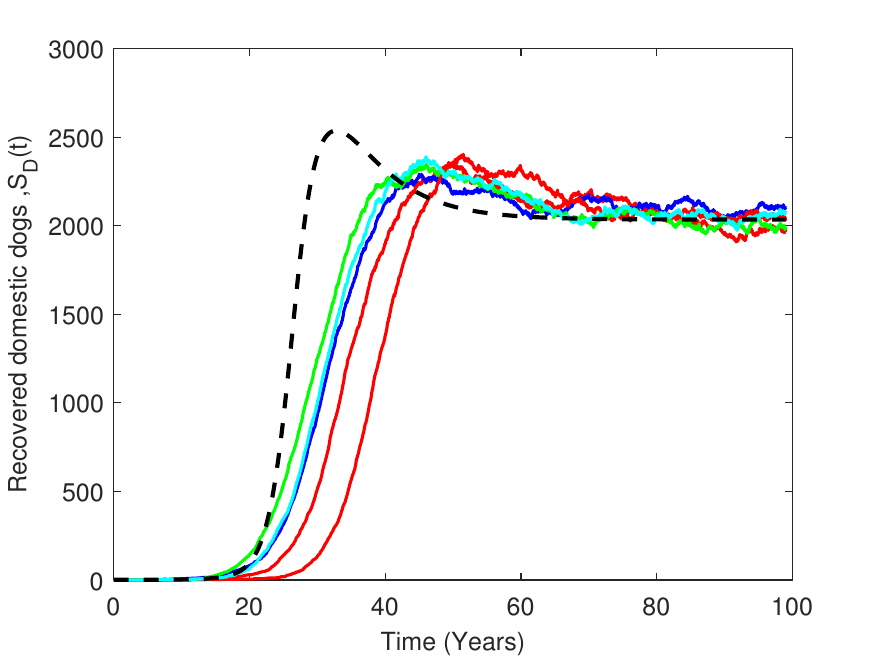}
\centering{(b)}
\end{minipage}
\centering  
\caption{\centering Comparison of Deterministic (dotted lines) 
and CTMC Sample Paths for Rabies Transmission Dynamics: 
(a)~Infected domestic dogs and (b)~Recovered domestic dogs.}
\label{Fig10}
\end{figure}

\begin{figure}[H]
\begin{minipage}[b]{0.45\textwidth}
\includegraphics[height=5.0cm, width=7.5cm]{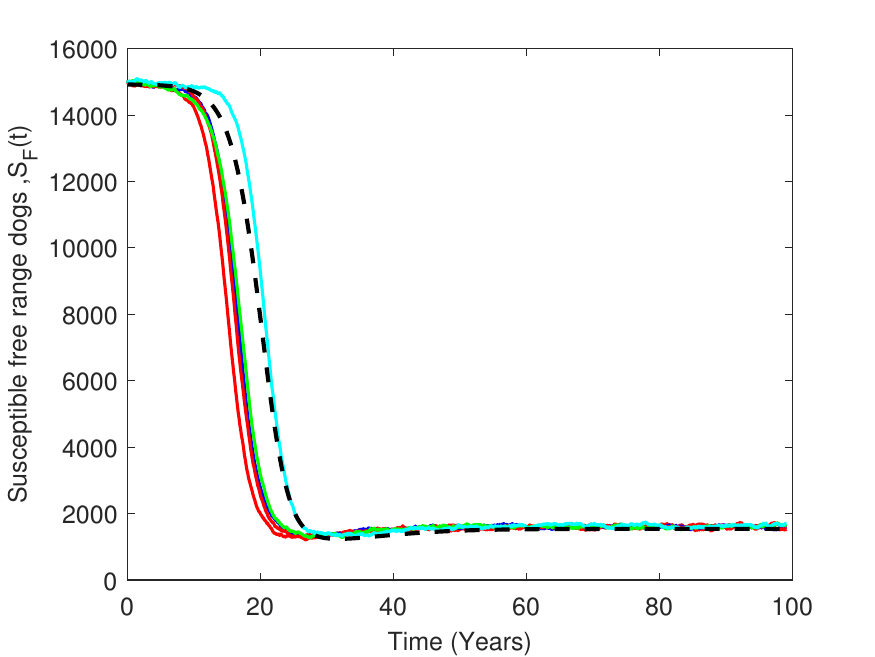}
\centering{(a)}
\end{minipage}
\begin{minipage}[b]{0.45\textwidth}
\includegraphics[height=5.0cm, width=7.5cm]{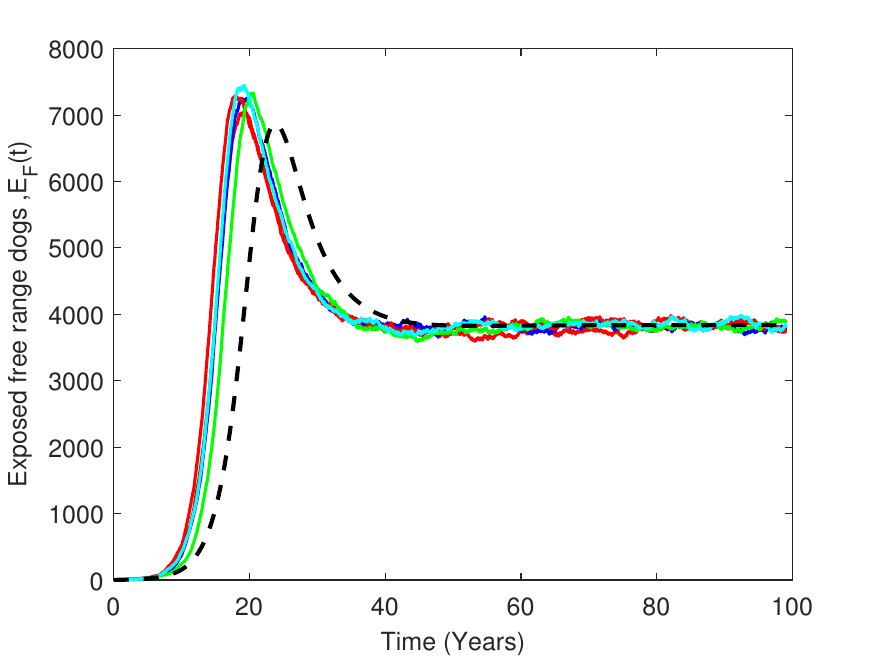}
\centering{(b)}
\end{minipage}
\begin{minipage}[b]{0.45\textwidth}
\includegraphics[height=5.0cm, width=7.5cm]{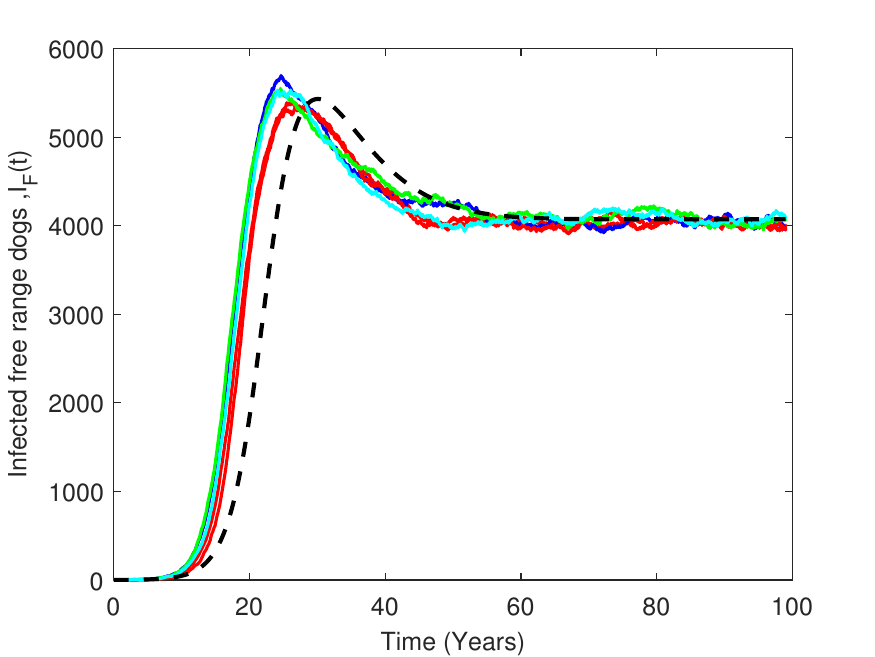}
\centering{(c)}
\end{minipage}
\centering  
\caption{\centering Comparison of Deterministic (dotted lines) 
and CTMC Sample Paths for Rabies Transmission Dynamics: 
(a)~Susceptible free range dogs  (b)~Exposed free range dogs 
and (c)~Infected free range dogs.}
\label{Fig11}
\end{figure}

Figures~\ref{Fig7}--\ref{Fig11} present both deterministic and continuous-time Markov chain (CTMC) 
stochastic results, revealing a comparable trend in rabies transmission dynamics. These figures 
demonstrate a reduction in susceptible populations following exposure, infection, and recovery 
events, with stabilization occurring after approximately 20 to 40 years. Likewise, susceptible 
humans, free-range dogs, and domestic dogs also experience a decline, ultimately reaching 
a steady state. Both modeling approaches exhibit a similar general pattern; deterministic 
results indicate an average trend across CTMC sample paths, while stochastic outputs 
reflect natural fluctuations. The relationship between the susceptible groups 
and the exposed, infected, and recovered classes is inversely related. Initially, 
the populations of exposed, infected, and recovered individuals both human and dogs 
see a rise, peaking around the first 20 years, followed by a gradual decline that 
stabilizes by year 30. This pattern suggests that the early increases in infections 
contribute to herd immunity within populations, ultimately leading to stabilization.


\section{Discussion and Conclusion}
\label{sec:5}

This study presents a comparative analysis of stochastic continuous-time 
Markov chains (CTMC) and deterministic models to understand rabies persistence 
in human and dog populations. Using the multitype branching process, 
the stochastic threshold for rabies persistence is established, offering 
new insights into how randomness affects disease extinction probabilities. 
Numerical simulations show that while the stochastic model outcomes closely 
align with deterministic results, stochasticity plays a key role in low-infection 
scenarios. Stochastic models help design flexible control strategies by accounting 
for uncertainties in disease spread, such as animal behavior or environmental factors 
(refer to Figures~\ref{Fig4}--\ref{Fig6}). Unlike deterministic models, 
which assume fixed rates, stochastic models adapt to varying scenarios, 
like regional differences or population behavior. These strategies include 
dynamic vaccination programs, real-time monitoring for adjustments, 
and focusing on high-risk areas for rabies transmission. Long-term planning, 
informed by continuous data, further refines interventions 
(refer to Figures~\ref{Fig7}--\ref{Fig10}).  
Our study provides a policy-driven perspective, advocating for a holistic 
rabies control approach by considering both predictable trends 
(deterministic models) and random events (stochastic models) 
(refer to Figures~\ref{Fig7}--\ref{Fig11}).


\section*{Data Availability}

The data used in this study are included in the manuscript.


\section*{Funding Statement}

Torres is supported by the Portuguese Foundation 
for Science and Technology (FCT) and CIDMA,
UID/4106/2025 and UID/PRR/4106/2025.


\section*{CRediT authorship contribution statement}

\textbf{Mfano Charles:} Writing -- original draft, Visualization, 
Validation, Software, Methodology, Formal analysis, Conceptualization.  

\textbf{Verdiana G. Masanja:} Writing -- review \& editing, 
Methodology, Formal analysis, Supervision.  

\textbf{Delfim F. M. Torres:} Writing -- review \& editing, 
Methodology, Formal analysis, Supervision.  

\textbf{Sayoki G. Mfinanga:} Writing -- review \& editing, Supervision.  

\textbf{G.A. Lyakurwa:} Writing -- review \& editing, Supervision.  


\section*{Conflicts of Interest}

The authors assert that they have no known conflicting financial interests 
or personal relationships that could have influenced 
the findings presented in this paper.


\section*{Acknowledgments}

The authors would like to acknowledge The Nelson Mandela African Institution 
of Science and Technology (NM-AIST) and the College of Business Education (CBE) 
for providing a conducive environment during the writing of this manuscript.
 

\appendix

\section{Global Stability of the Endemic Equilibrium Point $E^{*}$}
\label{Appendix:A}

Here we prove the result stated at the end of Section~\ref{sec:2.2.4}.

\begin{theorem}
The endemic equilibrium point  $E^{*}$ of the rabies model 
\eqref{eq:stochastic_model_with_noise} is globally asymptotically 
stable whenever $\mathcal{R}_{0}\ge1$.
\label{Th3}
\end{theorem}

\begin{proof}
To prove Theorem~\ref{Th3}, we adopt the approach of 
\cite{charles2024mathematical,chapwanya2022environment} 
by constructing a Lyapunov function of the form
\begin{equation*}
{\cal H}=\sum^{n}_{i=1}\mathcal{D}_{i}\left(U_{i}-U_{i}^{*}+U_{i}^{*}
\ln \left(\frac{U_{i}^{*}}{U_{i}}\right)\right), \mathcal{D}_{i}>0 
\;\;\text{for} \;\; i=1,2,3,\ldots,n,
\end{equation*}
where $\mathcal{D}_{i}$ represents a positive constant that needs to be determined, 
$U_{i}$ stands for the population variable at compartment $i$, and 
$U^{*}_{i}$ denotes the equilibrium point of the rabies model at compartment $i$ 
for $i \in \{1,2,3,\ldots,12\}$. Therefore, we define the Lyapunov $\mathcal{H}$ 
for model system \eqref{eq:stochastic_model_with_noise} as follows:
\begin{equation}
\mathcal{H}=
\begin{cases}
 \mathcal{D}_{1}\left(S_{H}-S_{H}^{*}+S_{H}\ln\left(\frac{S_{H}^{*}}{S_{H}}\right)\right)
+\mathcal{D}_{2}\left(E_{H}-E_{H}^{*}+E_{H}\ln\left(\frac{E_{H}^{*}}{E_{H}}\right)\right)
+\mathcal{D}_{3}\left(I_{H}-I_{H}^{*}+I_{H}\ln\left(\frac{I_{H}^{*}}{I_{H}}\right)\right)\\
+\mathcal{D}_{4}\left(R_{H}-R_{H}^{*}+R_{H}\ln\left(\frac{R_{H}^{*}}{R_{H}}\right)\right)
+\mathcal{D}_{5}\left(S_{F}-S_{F}^{*}+S_{F}\ln\left(\frac{S_{F}^{*}}{S_{F}}\right)\right)
+\mathcal{D}_{6}\left(E_{F}-E_{F}^{*}+E_{F}\ln\left(\frac{E_{F}^{*}}{E_{F}}\right)\right)\\
+\mathcal{D}_{7}\left(I_{F}-I_{F}^{*}+I_{F}\ln\left(\frac{I_{F}^{*}}{I_{F}}\right)\right)
+\mathcal{D}_{8}\left(S_{D}-S_{D}^{*}+S_{D}\ln\left(\frac{S_{D}^{*}}{S_{D}}\right)\right)
+\mathcal{D}_{9}\left(E_{D}-E_{D}^{*}+E_{D}\ln\left(\frac{E_{D}^{*}}{E_{D}}\right)\right)\\
+\mathcal{D}_{10}\left(I_{D}-I_{D}^{*}+I_{D}\ln\left(\frac{I_{D}^{*}}{I_{D}}\right)\right)
+\mathcal{D}_{11}\left(R_{D}-R_{D}^{*}+R_{D}\ln\left(\frac{R_{D}^{*}}{R_{D}}\right)\right)
+\mathcal{D}_{12}\left(M-M^{*}+M\ln\left(\frac{M^{*}}{M}\right)\right).
\end{cases}
\label{eqn14}
\end{equation}
Evaluating  equation $\left(\ref{eqn14}\right)$ at the
endemic equilibrium point $E^{*}$ gives 
\begin{equation*}
\mathcal{H} =\mathbb{E} ^{*}\left(S_{H}^{*}\;, E_{H}^{*}\;, I_{H}^{*}\;, 
R_{H}^{*} \;, S_{F}^{*} \;,E_{F}^{*}\;, I_{F}^{*}\;, S_{D}^{*}\;,
E_{D}^{*} \;,I_{D}^{*} \;, R_{D}^{*},M^{*}\right)=0.
\end{equation*}
Then, using the time derivative of the Lyapunov function \(\mathcal{H}\) 
in equation $\left(\ref{eqn14}\right)$ gives
\begin{equation}
\dfrac{d\mathcal{H}}{dt} 
\begin{cases}
&= \mathcal{D}_{1}\left(1-\dfrac{S^{*}_H}{S_{H}} \right)\dfrac{dS_H}{dt}
+\mathcal{D}_{2}\left(1-\dfrac{E^{*}_H}{E_H}\right)\dfrac{dE_H}{dt}
+\mathcal{D}_{3}\left(1-\dfrac{I^{*}_H}{I_H}\right)\dfrac{dI_H}{dt}
+\mathcal{D}_{4}\left(1-\dfrac{R^{*}_H}{R_H}\right)\dfrac{dR_H}{dt}\\ 
&+\mathcal{D}_{5}\left(1-\dfrac{S^{*}_F}{S_{F}} \right)\dfrac{dS_F}{dt}
+\mathcal{D}_{6}\left(1-\dfrac{E^{*}_F}{E_{F}} \right)\dfrac{dE_F}{dt}
+\mathcal{D}_{7}\left(1-\dfrac{I^{*}_F}{I_{F}} \right)\dfrac{dI_F}{dt}
+\mathcal{D}_{8}\left(1-\dfrac{S^{*}_D}{S_{D}} \right)\dfrac{dS_D}{dt}\\
&+\mathcal{D}_{9}\left(1-\dfrac{E^{*}_D}{E_{D}} \right)\dfrac{dE_D}{dt}
+\mathcal{D}_{10}\left(1-\dfrac{I^{*}_D}{I_{D}} \right)\dfrac{dI_D}{dt}
+\mathcal{D}_{11}\left(1-\dfrac{R^{*}_D}{R_{D}} \right)\dfrac{dR_D}{dt}
+\mathcal{D}_{12}\left(1-\dfrac{M^{*}}{M} \right)\dfrac{dM}{dt}.
\end{cases}
\label{eqn15}
\end{equation}
Consider the endemic equilibrium point \(E^{*}\) 
of equation \eqref{eq:stochastic_model_with_noise}such that 
\begin{equation}
\begin{cases}
\theta_{1} = \left(\tau_{1}I^{*}_{F}+\tau_{2}I^{*}_{D} + \tau_{3}
\lambda\left(M^{*}\right)\right)S^{*}_{H} + \mu_{1}S^{*}_{H} 
- \beta_{3}R^{*}_{H}, \;\; \mu_{1} + \beta_{1} + \beta_{2} 
= \dfrac{\left(\tau_{1}I^{*}_{F}+\tau_{2}I^{*}_{D}+\tau_{3}
\lambda\left(M^{*}\right)\right)S^{*}_{H}}{E^{*}_{H}}, \\ 
\sigma_{1}+\mu_{1} = \dfrac{\beta_{1}E^{*}_{H}}{I^{*}_{H}}, \;\; 
\beta_3+\mu_{1} = \dfrac{\beta_{2}E^{*}_{H}}{R^{*}_{H}}, \;\;
\theta_{2} = \left(\kappa_{1}I^{*}_{F}+\kappa_{2}I^{*}_{D} 
+ \kappa_{3}\lambda\left(M^{*}\right)\right)S^{*}_{F} + \mu_{2}S_{F}, \\
\mu_{2}+\gamma = \dfrac{\left(\kappa_{1}I^{*}_{F}+\kappa_{2}I^{*}_{D}
+\tau_{3}\lambda\left(M^{*}\right)\right)S^{*}_{F}}{E^{*}_{F}}, \;\; 
\sigma_{2}+\mu_{2} = \dfrac{\gamma E^{*}_{F}}{I^{*}_{F}}, \\ 
\theta_{3} = \left(\dfrac{\psi_1I^{*}_{F}}{1+\rho_{1}} 
+ \dfrac{\psi_2I^{*}_{D}}{1+\rho_{2}} + \dfrac{\psi_3
\lambda\left(M^{*}\right)}{1+\rho_{3}}\right)S^{*}_{D} 
+ \mu_{3}S^{*}_{D} - \gamma_{3}R^{*}_{D}, \;\; \mu_{3}
+\gamma_{1}+\gamma_{2} = \dfrac{\left(\dfrac{\psi_1I^{*}_{F}}{1+\rho_{1}}
+\dfrac{\psi_2I^{*}_{D}}{1+\rho_{2}}+\dfrac{\psi_3
\lambda\left(M^{*}\right)}{1+\rho_{3}}\right)S^{*}_{D}}{E^{*}_{D}}, \\
\sigma_{3}+\mu_{3} = \dfrac{\gamma_{1} E^{*}_{D}}{I^{*}_{D}}, \;\; 
\gamma_3+\mu_{3} = \dfrac{\gamma_{2}E^{*}_{D}}{R^{*}_{D}}, \;\; \mu_{4} 
= \dfrac{\left(\nu_{1}I^{*}_{H}+\nu_{2}I^{*}_{F}+\nu_{3}I^{*}_{D}\right)}{M^{*}}.
\end{cases}
\label{eqn16}
\end{equation}
Then, by substituting  \eqref{eqn16} 
into \eqref{eq:stochastic_model_with_noise}, we have
\begin{equation}
\dfrac{d\mathcal{H}}{dt}
= \begin{cases}
&\mathcal{D}_{1}\left(1-\frac{S^{*}_H}{S_{H}} \right)\left(\theta_{1}
+\beta_{3}R_{H}-\mu_{1} S_{H}-\chi_{1}\right)
+\mathcal{D}_{2}\left(1-\frac{E^{*}_H}{E_H} \right)\left(\chi_{1}
-\left(\mu_{1}+\beta_{1}+\beta_{2}\right)E_{H}\right)\\
&+\mathcal{D}_{3}\left(1-\frac{I^{*}_H}{I_H}\right)\left(\beta_{1}
E_{H}-\left(\sigma_{1}+\mu_{1}\right) I_{H}\right)
+\mathcal{D}_{4}\left(1-\frac{R^{*}_H}{R_H}\right)\left(\beta_{2} E_{H}
-\left(\beta_{3}+\mu_{1} \right) R_{H}\right)\\ 
&+\mathcal{D}_{5}\left(1-\frac{S^{*}_F}{S_{F}} \right)\left(\theta_{2}
-\chi_{2}-\mu_{2}S_{F}\right)+G_{6}\left(1-\frac{E^{*}_F}{E_{F}} 
\right)\left(\chi_{2}-\left(\mu_{2}+\gamma\right)E_{F}\right)\\
&+\mathcal{D}_{7}\left(1-\frac{I^{*}_F}{I_{F}} \right)\left(\gamma E_{F}
-\left(\mu_{2}+\sigma_{2}\right)I_{F}\right)
+\mathcal{D}_{8}\left(1-\frac{S^{*}_D}{S_{D}} \right)\left(\theta_{3}
-\mu_{3}S_{D}-\chi_{3}+\gamma_{3}R_{D}\right)\\
&+\mathcal{D}_{9}\left(1-\frac{E^{*}_D}{E_{D}} \right)\left(\chi_{3}
-\left(\mu_{3}+\gamma_{1}+\gamma_{2}\right) E_{D}\right)
+\mathcal{D}_{10}\left(1-\frac{I^{*}_D}{I_{D}} \right)\left(\gamma_{1}E_{D}
-\left(\mu_{3}+\delta_{3}\right) I_{D}\right)\\
&+\mathcal{D}_{11}\left(1-\frac{R^{*}_D}{R_{D}} \right)\left(\gamma_{2}
E_{D}-\left(\mu_{3}+\gamma_{3}\right)R_{D}\right)
+\mathcal{D}_{12}\left(1-\frac{M^{*}}{M} \right)\left(\left(\nu_1I_H
+\nu_2I_F+\nu_3I_D\right)-\mu_4M\right).
\end{cases}
\label{eqn27}
\end{equation}
Using the endemic equilibrium point  \(E^{0}\)
described in equation $\left(\ref{eqn16}\right)$,  
we simplify the equation $\left(\ref{eqn27}\right)$ as 
\begin{equation}
\frac{d\mathcal{H}}{dt}= 
\begin{cases}
&= \mathcal{D}_{1}\left(1-\frac{S^{*}_H}{S_{H}}\right)\Bigg(\left(\tau_{1}I^{*}_{F}
+\tau_{2}I^{*}_{D}+\tau_{3}\lambda\left(M^{*}\right)\right)S^{*}_{H}
+\mu_{1}S^{*}_{H}-\beta_{3}R^{*}_{H}-\mu_{1}S_{H}\\
&\quad -\left(\tau_{1}I_{F}+\tau_{2}I_{D}+\tau_{3}
\lambda\left(M\right)\right)S_{H}+\beta_{3}R_{H}\Bigg)\\
&+\mathcal{D}_{2}\left(1-\frac{E^{*}_H}{E_{H}}\right)\Bigg(
\left(\tau_{1}I_{F}+\tau_{2}I_{D}+\tau_{3}\lambda\left(M\right)\right)S_{H}
-\dfrac{\left(\tau_{1}I^{*}_{F}+\tau_{2}I^{*}_{D}+\tau_{3}
\lambda\left(M^{*}\right)\right)S^{*}_{H}E_{H}}{E^{*}_{H}}\Bigg)\\
&+\mathcal{D}_{3}\left(1-\frac{I^{*}_H}{I_{H}}\right)\left(\beta_{1}E_{H}
-\dfrac{\beta_{1}E^{*}_{H}I_{H}}{I^{*}_{H}}\right)
+\mathcal{D}_{4}\left(1-\frac{R^{*}_H}{R_{H}}\right)\left(\beta_{2}E_{H}
-\dfrac{\beta_{2}E^{*}_{H}R_{H}}{R^{*}_{H}}\right)\\
&+\mathcal{D}_{5}\left(1-\frac{S^{*}_F}{S_{F}}\right)\Bigg(\left(
\kappa_{1}I^{*}_{F}+\kappa_{2}I^{*}_{D}+\kappa_{3}
\lambda\left(M^{*}\right)\right)S^{*}_{F}+\mu_{2}S^{*}_{F}-\mu_{2}S_{F}\\
&\quad -\left(\kappa_{1}I_{F}+\kappa_{2}I_{D}
+\kappa_{3}\lambda\left(M\right)\right)S_{F}\Bigg)\\
&+\mathcal{D}_{6}\left(1-\frac{E^{*}_F}{E_{F}}\right)\Bigg(
\left(\kappa_{1}I_{F}+\kappa_{2}I_{D}+\kappa_{3}
\lambda\left(M\right)\right)S_{F}-\dfrac{\left(\tau_{1}I^{*}_{F}
+\kappa_{2}I^{*}_{D}+\kappa_{3}\lambda\left(M^{*}\right)\right)
S^{*}_{F}E_{F}}{E^{*}_{F}}\Bigg)\\
&+\mathcal{D}_{7}\left(1-\frac{I^{*}_F}{I_{F}} \right)\left(\gamma E_{F}
-\dfrac{\gamma E^{*}_{F}I_{F}}{I^{*}_{F}}\right)\\
&+\mathcal{D}_{8}\left(1-\frac{S^{*}_D}{S_{D}}\right)\Bigg(
\left(\dfrac{\psi_1I^{*}_{F}}{1+\rho_{1}}+\dfrac{\psi_2I^{*}_{D}}{1
+\rho_{2}}\dfrac{\psi_3\lambda\left(M^{*}\right)}{1+\rho_{3}}\right)
S^{*}_{D}+\mu_{3}S^{*}_{D}-\gamma_{3}R^{*}_{D}\\
&\quad -\left(\dfrac{\psi_1I_{F}}{1+\rho_{1}}+\dfrac{\psi_2I_{D}}{1+\rho_{2}}
\dfrac{\psi_3\lambda\left(M\right)}{1+\rho_{3}}\right)S_{D}
-\mu_{3}S_{D}+\gamma_{3}R_{D}\Bigg)\\
&+\mathcal{D}_{9}\left(1-\frac{E^{*}_D}{E_{D}}\right)\Bigg(\left(\dfrac{\psi_1I_{F}}{1
+\rho_{1}}+\dfrac{\psi_2I_{D}}{1+\rho_{2}}\dfrac{\psi_3\lambda\left(M\right)}{1
+\rho_{3}}\right)S_{D}-\dfrac{\left(\dfrac{\psi_1I^{*}_{F}}{1+\rho_{1}}
+\dfrac{\psi_2I^{*}_{D}}{1+\rho_{2}}\dfrac{\psi_3\lambda\left(M^{*}\right)}{1
+\rho_{3}}\right)S^{*}_{D}E_{D}}{E^{*}_{D}}\Bigg)\\
&+\mathcal{D}_{10}\left(1-\frac{I^{*}_D}{I_{D}}\right)\left(\gamma_{1}E_{D}
-\dfrac{\gamma_{1}E^{*}_{D}I_{D}}{I^{*}_{D}}\right)+\mathcal{D}_{11}\left(1
-\frac{R^{*}_D}{R_{D}} \right)\left(\gamma_{2}E_{D}
-\dfrac{\gamma_{2}E^{*}_{D}R_{D}}{R^{*}_{D}}\right)\\
&+\mathcal{D}_{12}\left(1-\frac{M^{*}}{M}\right)\left(\nu_{1}I_{H}+\nu_{2}I_{F}
+\nu_{3}I_{D}-\dfrac{\left(\nu_{1}I^{*}_{H}+\nu_{2}I^{*}_{F}
+\nu_{3}I^{*}_{D}\right)M}{M^{*}}\right).
\end{cases}
\label{eqn21}
\end{equation}
Then, equation \eqref{eqn21} can be expressed as follows: 
\begin{equation}
\dfrac{d\mathcal{H}}{dt} = 
\begin{cases}
-\mathcal{D}_{1}\mu_{1}S_{H}\left(1
-\dfrac{S^{*}_H}{S_{H}} \right)^{2}+\mathcal{D}_{1}\tau_{1}S_{H}
I_{F}\left(1-\dfrac{S^{*}_H}{S_{H}} \right)\left(
\dfrac{I^{*}_{F}S^{*}_{H}}{I_{F}S_{H}}-1\right)
+\mathcal{D}_{1}\tau_{2}S_{H}I_{D}\left(1-\dfrac{S^{*}_H}{S_{H}}\right)
\left(\dfrac{I^{*}_{D}S^{*}_{H}}{I_{D}S_{H}}-1\right)\\
+\mathcal{D}_{1}\tau_{3}S_{H}\lambda\left(M\right)\left(1-\dfrac{S^{*}_H}{S_{H}}\right)
\left(\dfrac{\lambda\left(M^{*}\right)S^{*}_{H}}{\lambda
\left(M\right)S_{H}}-1\right)+\mathcal{D}_{1}\beta_{3}R_{H}\left(1-\frac{S^{*}_H}{S_{H}}\right)
\left(1-\frac{R^{*}_H}{R_{H}} \right)\\
+\mathcal{D}_{2}\tau_{1}S_{H}I_{F}\left(1-\frac{E^{*}_H}{E_{H}}\right)
\left(1-\dfrac{I^{*}_{F}S^{*}_{H}E_{H}}{I_{F}S_{H}E^{*}_{H}}\right)
+\mathcal{D}_{2}\tau_{2}S_{H}I_{D}\left(1-\frac{E^{*}_H}{E_{H}}\right)
\left(1-\dfrac{I^{*}_{D}S^{*}_{H}E_{H}}{I_{D}S_{H}E^{*}_{H}}\right)\\
+\mathcal{D}_{2}\tau_{3}S_{H}\lambda\left(M\right)\left(1-\frac{E^{*}_H}{E_{H}}\right)
\left(1-\dfrac{\lambda\left(M^{*}\right)S^{*}_{H}E_{H}}{
\lambda\left(M\right)S_{H}E^{*}_{H}}\right)\\
+\mathcal{D}_{3}\beta_{1}E_{H}\left(1-\frac{I^{*}_H}{I_{H}}\right)
\left(1-\dfrac{E^{*}_{H}I_{H}}{E_{H}I^{*}_{H}}\right)
+\mathcal{D}_{4}\beta_{2}E_{H}\left(1-\frac{R^{*}_H}{R_{H}}\right)
\left(-\dfrac{E^{*}_{H}R_{H}}{E_{H}R^{*}_{H}}\right)\\
-\mathcal{D}_{5}\mu_{2}S_{F}\left(1-\frac{S^{*}_F}{S_{F}} \right)^{2}
+\mathcal{D}_{5}\kappa_{1}S_{F}I_{F}\left(1-\frac{S^{*}_F}{S_{F}}\right)
\left(\dfrac{I^{*}_{F}S^{*}_{F}}{I_{F}S_{F}}-1\right)
+\mathcal{D}_{5}\kappa_{2}S_{F}I_{D}\left(1-\frac{S^{*}_F}{S_{F}}\right)
\left(\dfrac{I^{*}_{D}S^{*}_{F}}{I_{D}S_{F}}-1\right)\\
+\mathcal{D}_{5}\kappa_{3}S_{F}\lambda\left(M\right)\left(1-\frac{S^{*}_{F}}{S_{F}}\right)
\left(\dfrac{\lambda\left(M^{*}\right)S^{*}_{F}}{\lambda\left(M\right)S_{F}}-1\right)
+ \mathcal{D}_{6}\kappa_{1}S_{F}I_{F}\left(1-\frac{E^{*}_F}{E_{F}}\right)
\left(1-\dfrac{I^{*}_{F}S^{*}_{F}E_{F}}{I_{F}S_{F}E^{*}_{F}}\right)\\
+\mathcal{D}_{6}\kappa_{2}S_{F}I_{D}\left(1-\frac{E^{*}_F}{E_{F}}\right)
\left(1-\dfrac{I^{*}_{D}S^{*}_{F}E_{F}}{I_{D}S_{F}E^{*}_{F}}\right)\\
+\mathcal{D}_{6}\kappa_{3}S_{F}\lambda\left(M\right)\left(1-\frac{E^{*}_F}{E_{F}}\right)
\left(1-\dfrac{\lambda\left(M^{*}\right)S^{*}_{F}E_{F}}{\lambda\left(M\right)
S_{F}E^{*}_{F}}\right)+\mathcal{D}_{7}\gamma E_{F}\left(1-\frac{I^{*}_F}{I_{F}}\right)
\left(1-\dfrac{E^{*}_{F}I_{F}}{E_{F}I^{*}_{F}}\right)\\
-\mathcal{D}_{8}\mu_{3}S_{D}\left(1-\frac{S^{*}_D}{S_{D}} \right)^{2}
+\dfrac{\psi_{1}S_{D}I_{F}\mathcal{D}_{8}}{\left(1+\rho_{1}\right)}
\left(1-\frac{S^{*}_D}{S_{D}}\right)\left(\dfrac{I^{*}_{F}
S^{*}_{D}}{I_{F}S_{D}}-1\right)+\dfrac{\psi_{2}S_{D}I_{F}
\mathcal{D}_{8}}{\left(1+\rho_{2}\right)}\left(1-\frac{S^{*}_D}{S_{D}}\right)
\left(\dfrac{I^{*}_{D}S^{*}_{D}}{I_{D}S_{D}}-1\right)\\
+\dfrac{\psi_{3}S_{D}\lambda\left(M\right)\mathcal{D}_{8}}{\left(1
+\rho_{3}\right)} \left(1-\frac{S^{*}_D}{S_{D}}\right)
\left(\dfrac{\lambda\left(M^{*}\right)S^{*}_{D}}{\lambda
\left(M\right)S_{D}}-1\right)+\mathcal{D}_{8}\gamma_{3}R_{D}\left(
1-\frac{S^{*}_D}{S_{D}} \right)\left(1-\frac{R^{*}_D}{R_{D}} \right)\\
+\dfrac{\psi_{1}S_{D}I_{F}\mathcal{D}_{9}}{\left(1+\rho_{1}\right)}
\left(1-\frac{E^{*}_D}{E_{D}} \right)\left(1-\dfrac{I^{*}_{F}S^{*}_{D}
E_{D}}{I_{F}S_{D}E^{*}_{D}}\right)+\dfrac{\psi_{2}S_{D}I_{F}
\mathcal{D}_{9}}{\left(1+\rho_{2}\right)}\left(1-\frac{E^{*}_D}{E_{D}}\right)
\left(1-\dfrac{I^{*}_{D}S^{*}_{D}E_{D}}{I_{D}S_{D}E^{*}_{D}}\right)\\
+\dfrac{\psi_{3}S_{D}I_{F}\mathcal{D}_{9}}{\left(1+\rho_{3}\right)}
\left(1-\frac{E^{*}_D}{E_{D}} \right)\left(1
-\dfrac{\lambda\left(M^{*}\right)S^{*}_{D}E_{D}}{\lambda
\left(M\right)S_{D}E^{*}_{D}}\right)\\
+\mathcal{D}_{10}\gamma_{1}E_{D}\left(1-\frac{I^{*}_D}{I_{D}}\right)
\left(1-\dfrac{E^{*}_{D}I_{D}}{E_{D}I^{*}_{D}}\right)
+\mathcal{D}_{11}\gamma_{2}E_{D}\left(1-\frac{R^{*}_D}{R_{D}}\right)
\left(-\dfrac{E^{*}_{D}R_{D}}{E_{D}R^{*}_{D}}\right)\\
+\mathcal{D}_{12}\nu_{1}I_{H}\left(1-\frac{M^{*}}{M} \right)
\left(1-\frac{I^{*}_{H}M}{I_{H}M^{*}} \right)
+\mathcal{D}_{12}\nu_{2}I_{F}\left(1-\frac{M^{*}}{M} \right)
\left(1-\frac{I^{*}_{F}M}{I_{F}M^{*}} \right)
+\mathcal{D}_{12}\nu_{3}I_{D}\left(1-\frac{M^{*}}{M} \right)
\left(1-\frac{I^{*}_{D}M}{I_{D}M^{*}} \right).
\end{cases}
\label{eqn23}
\end{equation}
Equation \eqref{eqn23} can be written as 
\begin{equation*}
\begin{array}{llll}
\dfrac{d\mathcal{H}}{dt}=\mathcal{G}+\mathcal{P},
\end{array}
\end{equation*}
where  
\begin{equation*}
\begin{array}{llll}
{\cal P}=-\mathcal{D}_{1}\mu_{1} S_{H}\left(1-\dfrac{S^{*}_H}{S_H}\right)^{2}
-\mathcal{D}_{5}\mu_{2} S_{F}\left(1-\dfrac{S^{*}_F}{S_F}\right)^{2}
-\mathcal{D}_{8}\mu_{3} S_{D}\left(1-\dfrac{S^{*}_D}{S_D}\right)^{2}
\end{array}
\end{equation*} 
and  
\begin{equation}
\mathcal{G} = 
\begin{cases}
\mathcal{D}_{1}\tau_{1}S_{H}I_{F}\left(1-\frac{S^{*}_H}{S_{H}} 
\right)\left(\dfrac{I^{*}_{F}S^{*}_{H}}{I_{F}S_{H}}-1\right)
+\mathcal{D}_{1}\tau_{2}S_{H}I_{D}\left(1-\frac{S^{*}_H}{S_{H}} \right)
\left(\dfrac{I^{*}_{D}S^{*}_{H}}{I_{D}S_{H}}-1\right)\\
+\mathcal{D}_{1}\tau_{3}S_{H}\lambda\left(M\right)\left(1-\frac{S^{*}_H}{S_{H}} \right)
\left(\dfrac{\lambda\left(M^{*}\right)S^{*}_{H}}
{\lambda\left(M\right)S_{H}}-1\right)+\mathcal{D}_{1}\beta_{3}R_{H}
\left(1-\frac{S^{*}_H}{S_{H}} \right)\left(1-\frac{R^{*}_H}{R_{H}} \right)\\
+\mathcal{D}_{2}\tau_{1}S_{H}I_{F}\left(1-\frac{E^{*}_H}{E_{H}} \right)
\left(1-\dfrac{I^{*}_{F}S^{*}_{H}E_{H}}{I_{F}S_{H}E^{*}_{H}}\right)
+\mathcal{D}_{2}\tau_{2}S_{H}I_{D}\left(1-\frac{E^{*}_H}{E_{H}} \right)
\left(1-\dfrac{I^{*}_{D}S^{*}_{H}E_{H}}{I_{D}
S_{H}E^{*}_{H}}\right)\\
+\mathcal{D}_{2}\tau_{3}S_{H}
\lambda\left(M\right)\left(1-\frac{E^{*}_H}{E_{H}} \right)
\left(1-\dfrac{\lambda\left(M^{*}\right)S^{*}_{H}
E_{H}}{\lambda\left(M\right)S_{H}E^{*}_{H}}\right)\\
+\mathcal{D}_{3}\beta_{1}E_{H}\left(1-\frac{I^{*}_H}{I_{H}}\right)
\left(1-\dfrac{E^{*}_{H}I_{H}}{E_{H}I^{*}_{H}}\right)
+\mathcal{D}_{4}\beta_{2}E_{H}\left(1-\frac{R^{*}_H}{R_{H}}\right)
\left(-\dfrac{E^{*}_{H}R_{H}}{E_{H}R^{*}_{H}}\right)\\
+\mathcal{D}_{5}\kappa_{1}S_{F}I_{F}\left(1-\frac{S^{*}_F}{S_{F}} \right)
\left(\dfrac{I^{*}_{F}S^{*}_{F}}{I_{F}S_{F}}-1\right)
+\mathcal{D}_{5}\kappa_{2}S_{F}I_{D}\left(1-\frac{S^{*}_F}{S_{F}} \right)
\left(\dfrac{I^{*}_{D}S^{*}_{F}}{I_{D}S_{F}}-1\right)\\
+\mathcal{D}_{5}\kappa_{3}S_{F}\lambda\left(M\right)\left(1-\frac{S^{*}_{F}}{S_{F}}\right)
\left(\dfrac{\lambda\left(M^{*}\right)S^{*}_{F}}{\lambda\left(M\right)S_{F}}
-1\right)+ \mathcal{D}_{6}\kappa_{1}S_{F}I_{F}\left(1-\frac{E^{*}_F}{E_{F}} \right)
\left(1-\dfrac{I^{*}_{F}S^{*}_{F}E_{F}}{I_{F}S_{F}E^{*}_{F}}\right)\\
+\mathcal{D}_{6}\kappa_{2}S_{F}I_{D}\left(1-\frac{E^{*}_F}{E_{F}} \right)
\left(1-\dfrac{I^{*}_{D}S^{*}_{F}E_{F}}{I_{D}S_{F}E^{*}_{F}}\right)\\
+\mathcal{D}_{6}\kappa_{3}S_{F}\lambda\left(M\right)\left(1-\frac{E^{*}_F}{E_{F}}\right)
\left(1-\dfrac{\lambda\left(M^{*}\right)S^{*}_{F}
E_{F}}{\lambda\left(M\right)S_{F}E^{*}_{F}}\right)+\mathcal{D}_{7}\gamma
E_{F}\left(1-\frac{I^{*}_F}{I_{F}}\right)
\left(1-\dfrac{E^{*}_{F}I_{F}}{E_{F}I^{*}_{F}}\right)\\
+\dfrac{\psi_{1}S_{D}I_{F}\mathcal{D}_{8}}{\left(1+\rho_{1}\right)}
\left(1-\frac{S^{*}_D}{S_{D}}\right)\left(\dfrac{I^{*}_{F}S^{*}_{D}}{I_{F}S_{D}}
-1\right)+\dfrac{\psi_{2}S_{D}I_{F}\mathcal{D}_{8}}{\left(1+\rho_{2}\right)}
\left(1-\frac{S^{*}_D}{S_{D}}\right)\left(\dfrac{I^{*}_{D}
S^{*}_{D}}{I_{D}S_{D}}-1\right)\\
+\dfrac{\psi_{3}S_{D}\lambda\left(M\right)\mathcal{D}_{8}}{\left(1
+\rho_{3}\right)} \left(1-\frac{S^{*}_D}{S_{D}}\right)
\left(\dfrac{\lambda\left(M^{*}\right)S^{*}_{D}}{\lambda
\left(M\right)S_{D}}-1\right)+\mathcal{D}_{8}\gamma_{3}
R_{D}\left(1-\frac{S^{*}_D}{S_{D}} \right)\left(1-\frac{R^{*}_D}{R_{D}} \right)\\
+\dfrac{\psi_{1}S_{D}I_{F}\mathcal{D}_{9}}{\left(1
+\rho_{1}\right)}\left(1-\frac{E^{*}_D}{E_{D}}\right)
\left(1-\dfrac{I^{*}_{F}S^{*}_{D}E_{D}}{I_{F}S_{D}E^{*}_{D}}\right)
+\dfrac{\psi_{2}S_{D}I_{F}\mathcal{D}_{9}}{\left(1+\rho_{2}\right)}\left(1
-\frac{E^{*}_D}{E_{D}} \right)
\left(1-\dfrac{I^{*}_{D}S^{*}_{D}E_{D}}{I_{D}S_{D}E^{*}_{D}}\right)\\
+\dfrac{\psi_{3}S_{D}I_{F}\mathcal{D}_{9}}{\left(1+\rho_{3}\right)}
\left(1-\frac{E^{*}_D}{E_{D}} \right)\left(1-\dfrac{\lambda\left(M^{*}\right)S^{*}_{D}
E_{D}}{\lambda\left(M\right)S_{D}E^{*}_{D}}\right)\\
+\mathcal{D}_{10}\gamma_{1}E_{D}\left(1-\frac{I^{*}_D}{I_{D}}\right)
\left(1-\dfrac{E^{*}_{D}I_{D}}{E_{D}I^{*}_{D}}\right)
+\mathcal{D}_{11}\gamma_{2}E_{D}\left(1-\frac{R^{*}_D}{R_{D}}\right)
\left(-\dfrac{E^{*}_{D}R_{D}}{E_{D}R^{*}_{D}}\right)\\ \\
+\mathcal{D}_{12}\nu_{1}I_{H}\left(1-\frac{M^{*}}{M} \right)
\left(1-\frac{I^{*}_{H}M}{I_{H}M^{*}} \right)+\mathcal{D}_{12}\nu_{2}I_{F}
\left(1-\frac{M^{*}}{M} \right)\left(1-\frac{I^{*}_{F}M}{I_{F}
M^{*}} \right)+\mathcal{D}_{12}\nu_{3}I_{D}\left(1-\dfrac{M^{*}}{M} \right)
\left(1-\dfrac{I^{*}_{D}M}{I_{D}M^{*}} \right).
\end{cases}
\label{eqn33}
\end{equation}
To simplify \eqref{eqn33}, let
\begin{equation*}
\begin{aligned}
a = \dfrac{S_{H}}{S^{*}_{H}}, \;
b = \dfrac{E_{H}}{E^{*}_{H}}, \;
c = \dfrac{I_{H}}{I^{*}_{H}}, \;
d = \dfrac{R_{H}}{R^{*}_{H}}, \;
e = \dfrac{S_{F}}{S^{*}_{F}}, \;
f = \dfrac{E_{F}}{E^{*}_{F}}, \;
g = \dfrac{I_{F}}{I^{*}_{F}}, \\
h = \dfrac{S_{D}}{S^{*}_{D}}, \;
r = \dfrac{E_{D}}{E^{*}_{D}}, \;
n = \dfrac{I_{D}}{I^{*}_{D}}, \;
m=\dfrac{\lambda\left(M\right)}{\lambda\left(M^{*}\right)},\;
l = \dfrac{R_{D}}{R^{*}_{D}}, \;\text{and}\;
k = \dfrac{M}{M^{*}}.
\end{aligned}
\end{equation*}
We express the  equation $\left(\ref{eqn33}\right)$ as
\begin{equation}
\mathcal{Q} = 
\begin{cases} 
\tau_{1}S_{H}I_{F}\left(1-\dfrac{b}{ac}+\dfrac{1}{ac}
-\dfrac{1}{b} \right)+\tau_{2}S_{H}I_{D}\left(\dfrac{1}{an}-1+\dfrac{1}{a^{2}n}
+\dfrac{1}{a}\right)+\tau_{3}S_{H}\lambda\left(M\right)
\left(\dfrac{1}{am}-1+\dfrac{1}{a^{2}m}+\dfrac{1}{a}\right)\\
+ \beta_{3}R_{H}\left(1-\dfrac{1}{d}-\dfrac{1}{a} +\dfrac{1}{ad}\right)
+ \tau_{1}S_{H}I_{F}\left(1-\dfrac{b}{af}-\dfrac{1}{b} 
+\dfrac{1}{af}\right)+\tau_{2}S_{H}I_{D}\left(1-\dfrac{b}{an}-\dfrac{1}{b}
+\dfrac{1}{an}\right)\\+\tau_{3}S_{H}\lambda\left(M\right)
\left(1-\dfrac{b}{af}-\dfrac{1}{b} +\dfrac{1}{af}\right)
+\tau_{2}S_{H}I_{D}\left(1-\dfrac{b}{am}-\dfrac{1}{b} 
+\dfrac{1}{am}\right)+\beta_{1}E_{H}\left(1-\dfrac{b}{c}
-\dfrac{1}{c}+\dfrac{b}{c^{2}}\right)\\+\beta_{2}E_{H}
\left(1-\dfrac{b}{d}-\dfrac{1}{d}+\dfrac{b}{d^{2}}\right)
+\kappa_{1}S_{F}I_{F}\left(\dfrac{1}{ef}-1-\dfrac{1}{e^{2}f}
+\dfrac{1}{e}\right)+\kappa_{2}S_{F}I_{D}\left(\dfrac{1}{en}
-1-\dfrac{1}{e^{2}n}+\dfrac{1}{e}\right)\\
+\kappa_{3}S_{F}\lambda\left(M\right)\left(\dfrac{1}{em}-1
-\dfrac{1}{e^{2}m}+\dfrac{1}{e}\right)
+\kappa_{1}S_{F}I_{F}\left(1-\dfrac{f}{en}-\dfrac{1}{f}
+\dfrac{1}{en}\right)+\kappa_{2}S_{F}I_{D}\left(1-\dfrac{f}{en}
-\dfrac{1}{f}+\dfrac{1}{en}\right)\\+\kappa_{3}S_{F}\lambda
\left(M\right)\left(1-\dfrac{f}{me}-\dfrac{1}{f}+\dfrac{1}{me}\right)
+\gamma E_{F}\left(1-\dfrac{g}{f}-\dfrac{1}{f}+\dfrac{g}{f^{2}}\right)
+\dfrac{\psi_{1}S_{D}I_{F}}{\left(1+\rho_{1}\right)}\left(1-\dfrac{1}{h}
-\dfrac{1}{h^{2}g}+\dfrac{1}{h}\right)\\+\dfrac{\psi_{2}S_{D}I_{F}}{\left(1
+\rho_{2}\right)}\left(1-\dfrac{1}{h}-\dfrac{1}{h^{2}g}+\dfrac{1}{h}\right)
+\dfrac{\psi_{3}S_{D}\lambda\left(M\right)}{\left(1+\rho_{3}\right)}
\left(\dfrac{1}{mh}-1-\dfrac{1}{h^{2}m}+\dfrac{1}{h}\right)
+\gamma_{3}R_{D}\left(1-\dfrac{1}{l}-\dfrac{1}{h}+\dfrac{1}{hl}\right)\\
+\dfrac{\psi_{1}S_{D}I_{F}}{\left(1+\rho_{1}\right)}
\left(1-\dfrac{r}{hg}-\dfrac{1}{r}+\dfrac{1}{hg}\right)
+\dfrac{\psi_{2}S_{D}I_{F}}{\left(1+\rho_{2}\right)}
\left(1-\dfrac{r}{hn}-\dfrac{1}{r}+\dfrac{1}{hn}\right)
+\dfrac{\psi_{3}S_{D}I_{F}}{\left(1+\rho_{3}\right)}
\left(1-\dfrac{r}{hm}-\dfrac{1}{r}+\dfrac{1}{hm}\right)\\
+\gamma_{1}E_{D}\left(1-\dfrac{l}{r}-\dfrac{1}{l}+\dfrac{1}{r}\right)
+\gamma_{2}E_{D}\left(1-\dfrac{r}{hg}-\dfrac{1}{r}+\dfrac{1}{hg}\right)
+\nu_{1}I_{H}\left(1-\dfrac{k}{c}-\dfrac{1}{k}+\dfrac{1}{c}\right)
+\nu_{2}I_{F}\left(1-\dfrac{k}{g}-\dfrac{1}{k}+\dfrac{1}{g}\right)\\
+\nu_{3}I_{D}\left(1-\dfrac{k}{n}-\dfrac{1}{k}+\dfrac{1}{n}\right). 
\end{cases}
\label{eqn36}
\end{equation}
From equation $\left(\ref{eqn36}\right)$, we have
\begin{equation}
\begin{aligned}
1- \dfrac{1}{d}-\dfrac{1}{a}+\dfrac{1}{ad}
=\left(1-\dfrac{1}{d}\right)+\left(1-\dfrac{1}{a}\right)
-\left(1-\dfrac{1}{ad}\right).
\end{aligned}
\label{def2}
\end{equation}
To proceed, we make use of the following basic inequality:
if $\epsilon(y) = 1 - y + \ln y $, then $\epsilon(y) \leq 0$  
such  that  $1 - y \leq -\ln y$ if, and only if, $y > 0$
and, from the concept of geometric mean, 
equation \eqref{def2} is written as
\begin{equation}
\begin{split}
\left(1-\dfrac{1}{d}\right)+\left(1-\dfrac{1}{a}\right)
-\left(1-\dfrac{1}{ad}\right)
&\le -\ln\left(\dfrac{1}{d}\right)
-\ln\left(\dfrac{1}{a}\right)+\ln\left(\dfrac{1}{ad}\right)\\
&\le \ln\left(a\times d\times\dfrac{1}{ad} \right)=\ln\left(1\right)=0.
\end{split}
\label{def22}
\end{equation}
Following similar procedures as in \eqref{def22}, we get
\begin{equation*}
\begin{aligned}
1-\dfrac{c}{b}-\dfrac{1}{c}+\dfrac{1}{b}\le 0,
\quad 1-\dfrac{p}{m} -\dfrac{1}{p}+\dfrac{1}{m}\le 0,
\quad 1-\dfrac{d}{b}-\dfrac{1}{d}+\dfrac{1}{b} \le 0.
\end{aligned}
\end{equation*}
From equation \eqref{eqn23}, the global stability holds only if 
$\dfrac{d\mathcal{H}}{dt}\leq0$. Now, if $\mathcal{P}<\mathcal{G}$, then 
$\dfrac{d\mathcal{H}}{dt}$ will be negative definite, which implies that 
$\dfrac{d\mathcal{H}}{dt}<0$ and $\dfrac{d\mathcal{H}}{dt}=0$ only at the 
endemic equilibrium point $E^{*}$. Hence, by LaSalle's 
invariance principle \cite{lasalle1976stability}, any solution 
to the rabies model \eqref{eq:stochastic_model_with_noise} which intersects 
the interior $\mathbb{R}_{+}^{12}$ limits to $ E^{*}$ is globally asymptotically 
stable whatever ${\cal R}_0>1$.
\end{proof}
 


\end{document}